\documentclass[
    aps,
	twocolumn,
    groupeaddress,
	longbibliography,
]{revtex4-2}
\usepackage[english]{babel}
\usepackage[utf8]{inputenc}
\usepackage{siunitx}
\usepackage{soul}
\usepackage{lipsum}
\usepackage{rotating}
\usepackage{multirow} 
\usepackage{booktabs}
\usepackage{pifont}

\usepackage{ifthen}
\usepackage{tikz}
\usetikzlibrary{arrows}
\usetikzlibrary {arrows.meta}
\usetikzlibrary{arrows,decorations.pathmorphing}
\usetikzlibrary{calc}
\usepackage{empheq} 
\usepackage{amsmath,amssymb,amsfonts}
\usepackage{color}
\usepackage{xcolor}
\usepackage{graphicx}
\graphicspath{{Images/}}
\usepackage{subcaption}
\usepackage{placeins}
\usepackage{caption} 
\usepackage{physics}
\usepackage{float}
\usepackage[normalem]{ulem}
\usepackage{tabularx}
\usepackage{bm}
\usepackage{comment}

\renewcommand{\selectlanguage}[1]{} 

\usepackage[
	colorlinks=true,
	linkcolor=blue,
	urlcolor=blue,
	citecolor=blue
]{hyperref}

\emergencystretch=\maxdimen
\newcommand{\dtuelectro}{
    Department of Electrical and Photonics Engineering, Technical University of Denmark,
    2800 Kgs. Lyngby,
    Denmark
}

\begin{document}


\title{
Photonic ``hourglass" design beyond the standard bulk model of phonon decoherence
}

\author{José Ferreira Neto}
\email{jofer@dtu.dk}
\affiliation{\dtuelectro}

\author{Benedek Gaál}
\affiliation{\dtuelectro}

\author{Luca Vannucci}
\affiliation{\dtuelectro}

\author{Niels Gregersen}
\affiliation{\dtuelectro}

\date{\today}

\begin{abstract}

We study the impact of mechanical vibrations on the performance of the photonic ``hourglass" structure, which is predicted to emit single photons on-demand with near-unity efficiency and indistinguishability.
Previous investigations neglected the impact of vibrational modes inherent to this quasi-1D geometry, relying instead on a three-dimensional bulk assumption for the phonon modes.
However, it has been shown that phonon decoherence has a much stronger impact in 1D structures as compared with bulk media.
Here, we surprisingly demonstrate the robustness of the photonic hourglass design, achieving close-to-unity indistinguishability even by incorporating a detailed description of the vibrational modes.
We explain this unexpected result in terms of the large Purcell enhancement of the hourglass single-photon source, which eliminates the negative effect of phonons.
Our findings highlight the key role of high-$Q$ optical cavities in mitigating the detrimental effect of phonon decoherence, even for structures of reduced dimensionality.

\end{abstract}

\maketitle

\section{Introduction}


Within quantum physics, the pursuit of quantum computing has been a long-sought goal \cite{feynman_simulating_1982,noauthor_40_2022}. Among the various physical platforms proposed, the photonic quantum computer stands out as a particularly promising avenue \cite{politi_shors_2009,crespi_integrated_2011,humphreys_linear_2013,zhong_quantum_2020}. Here, a key component is the single-photon source (SPS) allowing for directional, triggered emission and efficient collection of highly pure indistinguishable single photons acting as carriers of quantum information. The figures of merit include \cite{piprek_handbook_2017,heindel_quantum_2023} the photon indistinguishability $\mathcal{I}$, which measures the degree to which photons emitted from the source are quantum-mechanically identical, reflecting their capacity to coalesce into a two-photon state in a Hong-Ou-Mandel experiment \cite{hong_measurement_1987}, and the photon collection efficiency $\mathcal{\varepsilon}$, defined as the probability that the collecting optics captures a photon emitted by the source. 
While highly indistinguishable photon emission can be achieved using spontaneous parametric down-conversion \cite{kwiat_new_1995}, this process is probabilistic, limiting the efficiency to a few percent. 

As an alternative, the semiconductor quantum dot (QD) positioned inside a tailored photonic environment has recently emerged as a successful strategy \cite{senellart_high-performance_2017,heindel_quantum_2023} for deterministic single-photon generation. Indeed, by placing the QD inside an optical cavity and exploiting cavity quantum electrodynamics (cQED), highly efficient emission $\mathcal{\varepsilon} \sim 0.6$ \cite{wang_towards_2019,tomm_bright_2021,maring_versatile_2024} of single indistinguishable photons has been demonstrated. However, the cQED approach suffers from an inherent trade-off \cite{iles-smith_phonon_2017} between the achievable efficiency and indistinguishability in the presence of phonon-induced decoherence, limiting the achievable product $\mathcal{I}\varepsilon$ for the micropillar SPS to $\sim$0.95 \cite{wang_micropillar_2020}. 

\begin{figure}[b]
\centering
\includegraphics[width=0.80\linewidth]{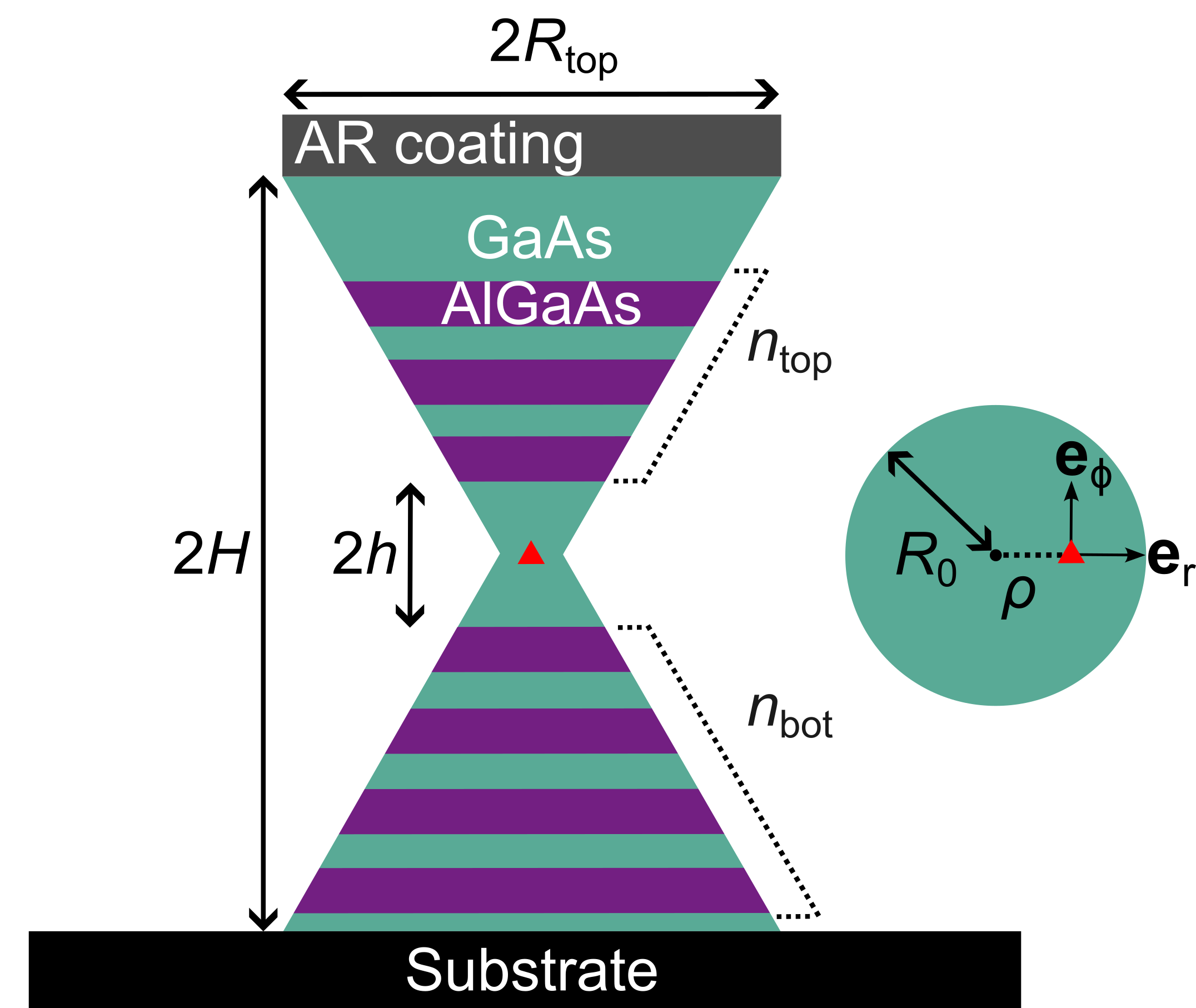}
\caption{The hourglass structure hosting a pyramidal QD (red triangle) and beam waist cross-section featuring an off-axis dipole displaced from the center by $\rho$.}
\label{fig:sketch}
\end{figure}

The hybrid photonic nanowire-micropillar hourglass design \cite{osterkryger_photonic_2019,gaal_near-unity_2022} sketched in Fig.\ \ref{fig:sketch} allows circumventing this trade-off by exploiting suppression of the background emission \cite{wang_suppression_2021}, and a product $\varepsilon\mathcal{I} \sim 0.973$  \cite{gaal_near-unity_2022} was predicted with a computed indistinguishability  $\mathcal{I}$ above 0.99. This prediction of the indistinguishability in the presence of phonon-induced decoherence was obtained using a Born-Markov polaron master equation for bulk media \cite{gaal_near-unity_2022,iles-smith_phonon_2017}. This approach models the dynamics of an open quantum system interacting with a memoryless phonon reservoir, where the interaction with the mechanical environment of the hourglass structure is assumed to be the same as for a 3D bulk structure. However, recent studies show \cite{tighineanu_phonon_2018} that low-dimensional structures, in particular the 1D wire geometry, feature significantly reduced  indistinguishability compared with a 3D bulk medium, thus questioning the validity of the mechanical model and the corresponding high indistinguishability reported in Ref.\ \cite{gaal_near-unity_2022}. 

In this work, we investigate the two-photon indistinguishability $\mathcal{I}$ of the photonic hourglass structure in the presence of temperature-induced dephasing via longitudinal acoustic phonons using a model of the phonon spectral density accounting for the discrete set of mechanical vibration modes of the hourglass structure. In particular, we elucidate whether or not the trade-off \cite{iles-smith_phonon_2017} between $\varepsilon$ and $\mathcal{I}$ can still be circumvented by the proper structuring of the photonic environment \cite{gaal_near-unity_2022} when fully taking into account the lattice vibrations characteristic of a low-dimensional geometry.

\begin{figure*}[t] 
\centering
\begin{subfigure}{\textwidth}
  \includegraphics[width=\linewidth]{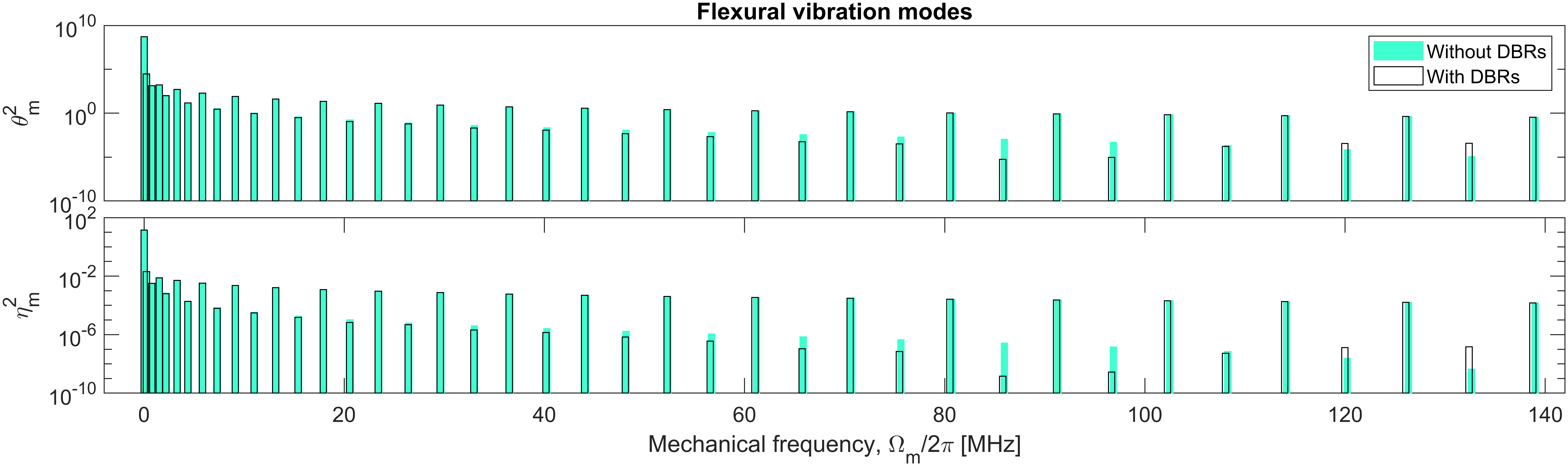}
\end{subfigure}%
\vspace{10pt}
\begin{subfigure}{\textwidth}
  \includegraphics[width=\linewidth]{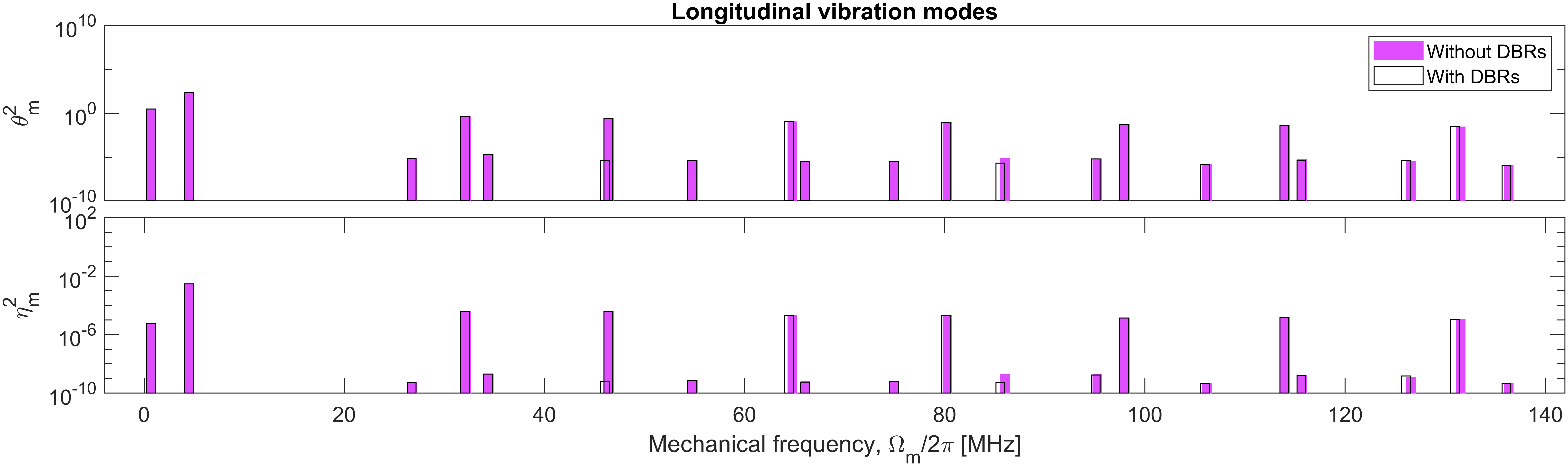}
\end{subfigure}
\caption{Examining the impact of thermally-induced vibrations on the photonic hourglass structure at a temperature of $T=\SI{4}{\kelvin}$, we present the parameters $\theta_{m}^{2}$ and $\eta_{m}^{2}$ for all considered vibration modes. For flexural modes, we illustrate the maximum, on-sidewall values. In contrast, we depict the on-axis results for longitudinal modes, as they contribute more significantly to the $\epsilon_{zz}$ strain component.}
\label{fig:final_vibs}
\end{figure*}

\section{Geometry}

The geometry of our photonic hourglass structure is illustrated in Fig.~\ref{fig:sketch} with geometric parameters almost identical, except for a larger number of distributed Bragg reflector (DBR) layer pairs in our bottom DBR, to those in Ref.\ \cite{gaal_near-unity_2022}. The cavity, formed from 11 (48) DBR pairs in the top (bottom) mirror,  is designed to operate at a wavelength $\lambda = 925\ \mathrm{nm}$, a typical emission wavelength for InAs/GaAs QDs. At the position of the QD, the  design features a narrow  diameter of $2R_{0} = 228\ \mathrm{nm}$ combined with a large top diameter $2R_{\mathrm{top}} = 1860\ \mathrm{nm}$. The total height of the structure $h_{\mathrm{total}} = 2H + t_{\mathrm{AR}} \sim 117$ \textmu $\mathrm{m}$, where $H = 58.5$ \textmu$\mathrm{m}$ and the thickness of the anti-reflection (AR) coating $t_{\mathrm{AR}} = \lambda/4n_{\mathrm{eff,AR}}$ \cite{gaal_near-unity_2022}. The exceptionally high aspect ratio of $\sim$513 is due to the very small sidewall angle required to ensure adiabatic mode expansion \cite{osterkryger_photonic_2019,gregersen_broadband_2016}. This aspect ratio greatly exceeds the one of a typical micropillar SPS \cite{wang_micropillar_2020,reitzenstein_quantum_2010,ding_-demand_2016}, suggesting that a 3D description of the vibrational modes may  indeed be inaccurate. Ultimately, the very high taper height of 58.5 {\textmu}m is a consequence of two fundamental requirements: (i) a sufficiently large top radius to ensure a Gaussian far-field profile and (ii) the aforementioned very small sidewall angle to yield near-unity transmission. One could potentially increase the beam waist diameter 2$R_{0}$ from 228 nm to 500 nm, while preserving the original sidewall angle. This would substantially reduce the aspect ratio of the geometry from 513 to 195. However, fabricating such a structure would still pose a significant challenge. While this potential improvement could be explored in future work by adopting nonlinear taper shapes, e.g., horn shape \cite{burns_optical_1977}, to further reduce the taper height and bridge the gap between theory and fabrication methods. We emphasize that the primary aim of this work is not to refine the optical design but rather to focus on a theoretical reevaluation, leaving advancements in optical design to follow-up studies. The refractive indices of the alternating DBR layers are $n_{\mathrm{AlGaAs}} = 2.9895$ and $n_{\mathrm{GaAs}} = 3.4788$, while $n_{\mathrm{AR}} = \sqrt{n_{\mathrm{GaAs}}}$. Additional material parameters are given in Table \textcolor{blue}{A}  in the supplementary material.

\section{Model}

Our primary goal is to evaluate the performance of the hourglass structure predicted to feature very high indistinguishability ($\mathcal{I} > 0.99$) under a 3D bulk phonon assumption. Thus, to distinguish between optical and mechanical contributions to the performance, we consider two distinct structures: (i) the optimized structure with DBRs, as depicted in Fig.~\ref{fig:sketch}, and (ii) the geometry without DBRs, where the entire hourglass geometry is instead homogeneously composed of GaAs material. 

Assuming a perfectly antibunched photon output and an initially excited emitter, we can calculate the two-photon indistinguishability $\mathcal{I}$ as \cite{artioli_design_2019}

\begin{equation}
\label{eqn:2_model}
\mathcal{I}=\Gamma\int_{0}^{\infty} \mathrm{d}t\ e^{-\Gamma t}|P(t)|^{2},
\end{equation}
where we consider the emission rate into the cavity $\Gamma=\Gamma_{\mathrm{C}}=F_{\mathrm{p}} \Gamma_{\mathrm{Bulk}}$, which is enhanced by the Purcell factor $F_{\mathrm{p}}$ relative to its value $\Gamma_{\mathrm{Bulk}}$ in a bulk medium, and where we assume $\Gamma_{\mathrm{Bulk}}$ = 1 GHz. 
As discussed in detail in Appendix \ref{section:appendix_A}, the QD coherence evolution is given in an independent boson model by $P(t)=e^{\Phi(t)}$, where the phonon propagator $\Phi(t)=\sum_{m}-\theta_{m}^{2}\sin^{2}(\omega_{m}t/2)-\eta_{m}^{2}(1-e^{-i\omega_{m}t})$. Here, $\omega_{m}$ are the mechanical eigenfrequencies with associated QD coupling parameters $\theta_{m}$ and $\eta_{m}$ defined in Appendix \ref{section:appendix_A}.

The modulation of the QD bandgap via deformation potential is given by \cite{pearson_analytical_2000,stepanov_large_2016}
\begin{equation}
\label{eqn:1_model}
\hbar\frac{\partial \omega_{\mathrm{eg}}}{\partial u_{m}}=a\left(\frac{\partial \epsilon_{\mathrm{h}}}{\partial u_{m}}\right)+\frac{b}{2}\left(\frac{\partial \epsilon_{\mathrm{sh}}}{\partial u_{m}}\right), 
\end{equation}
where our pyramidal QD experiences a residual bi-axial compressive strain given by the hydrostatic $\epsilon_{\mathrm{h}}=\epsilon_{xx}+\epsilon_{yy}+\epsilon_{zz}$ and the tetragonal shear $\epsilon_{\mathrm{sh}}=2\epsilon_{zz}-\epsilon_{xx}-\epsilon_{yy}$ ($z$ is the QD growth axis) strain contributions, with $a$ and $b$ corresponding to the associated deformation potentials of the QD material, respectively. Here, we suppose that the QD is composed of an $\mathrm{In}_{0.5}\mathrm{Ga}_{0.5}\mathrm{As}$ alloy and use $a = -7.5$ eV and $b = -1.9$ eV (linear interpolation between the deformation potentials on InAs and GaAs \cite{vurgaftman_band_2001}). 

The vibrational properties, including the mechanical eigenfrequencies $\omega_{m}$ for the free-standing geometry (singly-clamped to a substrate), their effective mass $m^{\mathrm{eff}}_{m}$ (see supplementary material for more details), and primary strain tensors ($\epsilon_{xx}$, $\epsilon_{yy}$, $\epsilon_{zz}$), were calculated with the finite element method (FEM) software \textsc{\footnotesize{COMSOL}} Multiphysics. The strain tensors are extracted at points along the beam waist cross-section defined by the direction of maximum displacement, i.e., at a height $H$ from the substrate, as shown in Fig.~\ref{fig:sketch}. The retrieved modes are then classified as flexural ($F$), longitudinal ($L$), or torsional ($T$) so that we can properly account for their contributions. It should be noted that the present method based on FEM simulations is well suited to study the characteristic low-frequency mechanical modes in the MHz--GHz regime, where the emitter can be considered as a pointlike source compared with the phonon wavelength \cite{artioli_design_2019}. The influence of high-frequency vibrational modes in the THz range has been considered in a separate work, where the structure is modeled as a semi-infinite 1D wire, and the finite size of the QD is taken into account \cite{ferreira_neto_one-dimensional_2024}. There, it was demonstrated that Purcell enhancement of the emission rate makes it possible to obtain near-unity indistinguishability despite the presence of a rich sideband extending to THz frequencies.

We perform an assessment of the coupling parameters $\eta_{m}^{2}=(g_{m}/\omega_{m})^{2}$ and $\theta_{m}^{2}=4\eta_{m}^{2}N_{m}$. These parameters indicate phonon emission in the zero-temperature limit and thermally-induced absorption, respectively. Here, $g_{m}$ corresponds to the optomechanical coupling strength and $N_{m}=[\exp(\hbar\omega_{m}/k_{\mathrm{B}}T)-1]^{-1}$ represents the phonon occupation number dictated by Bose-Einstein statistics. Among the first 101 modes analyzed (ranging up to 140 MHz), 80 exhibit flexural behavior (consisting of orthogonal flexural doublets \cite{braakman_force_2019}) while 21 are longitudinal. The upper panel of Fig.~\ref{fig:final_vibs} illustrates 40 flexural modes vibrating along the same axis, whereas the lower panel features modes with longitudinal behavior. By comparing the mechanical frequencies and their associated couplings for both structures, i.e., with and without DBRs, it becomes clear that they display very similar results from a purely \textit{mechanical} point of view. The exact values for the resonance frequencies of both flexural and longitudinal modes, along with their corresponding coupling strengths, are presented in Tables \textcolor{blue}{B} and \textcolor{blue}{C} in the supplementary material. As depicted in Fig.~\ref{fig:final_vibs}, low-frequency vibration modes are the primary source of dephasing, with the fundamental flexural mode $F_{1}$ being the most prominent, significantly exceeding the second flexural mode by more than four orders of magnitude. For $\theta^{2}_{m} \gg 1$, the QD experiences sizable spectral broadening, directly reflecting the thermally driven stress fluctuations. Conversely, as we move into the higher-frequency regime ($\theta^{2}_{m} \ll 1$), these contributions diminish substantially, leading merely to minor perturbative corrections. For the exact values of the extracted coupling parameters $\theta^{2}_{m}$ and $\eta^{2}_{m}$, check Tables \textcolor{blue}{D} and \textcolor{blue}{E} in the supplementary material.

We evaluate the Purcell factor $F_{\mathrm{p}}$ and the efficiency $\varepsilon$ using an open-geometry Fourier Modal Method (FMM) \cite{lavrinenko_numerical_2018,gur_open-geometry_2021}, where the QD at the position $\mathbf{r}_{0}$ is modeled as a point dipole $\mathbf{p}\delta(\mathbf{r}-\mathbf{r}_{0})$ with in-plane polarization \textbf{p}. 
The efficiency is computed using a single-mode model given by $\varepsilon = \beta \gamma$ \cite{gaal_near-unity_2022,gregersen_broadband_2016}. Here, 
the spontaneous emission $\beta$ factor describes the fraction of light emitted into the cavity  given by $\beta = \frac{\Gamma_{\mathrm{C}}}{\Gamma_{\mathrm{T}}} = \frac{\Gamma_{\mathrm{C}}}{\Gamma_{\mathrm{\mathrm{C}}}+\Gamma_{\mathrm{B}}}$, where $\Gamma_{\mathrm{T}} = \Gamma_{\mathrm{C}} + \Gamma_{\mathrm{B}}$ is the total emission rate including the contribution $\Gamma_{\mathrm{B}}$ to background modes. Additionally, the transmission $\gamma$ describes the power detected by a collection lens with specific numerical aperture (NA), taking into account an overlap with a Gaussian profile \cite{munsch_dielectric_2013,munsch_erratum_2013}, relative to the power in the cavity. 
This single-mode model provides an excellent description \cite{gaal_near-unity_2022} of the efficiency combined with physical insight into the governing physics. 

\begin{figure}[hbtp] 
  \includegraphics[width=\linewidth]{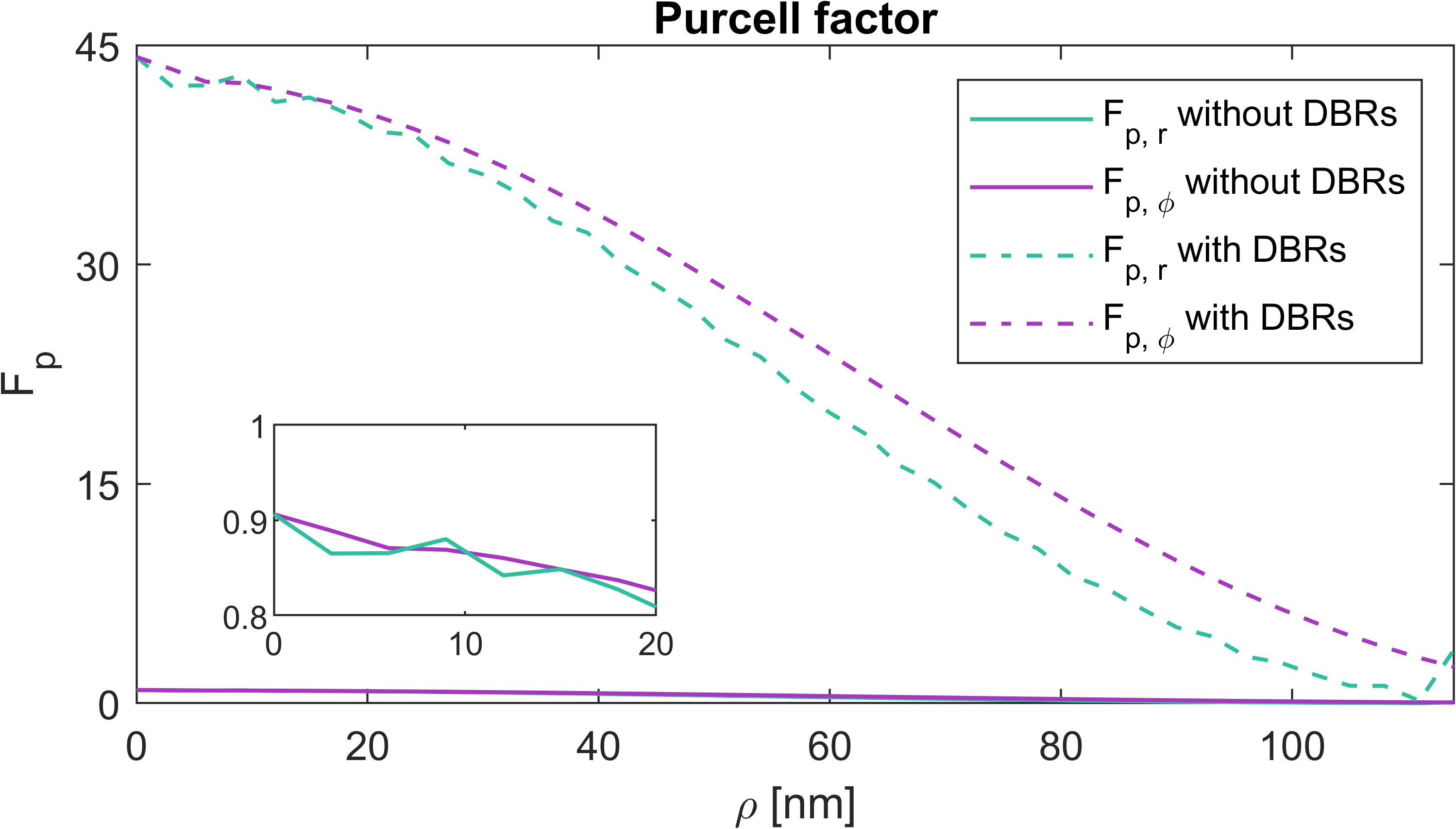}
\caption{Computed Purcell factor $F_{\mathrm{p}}$ for a dipole with a given polarization as a function of its radial displacement from the center. Here, the subscript $r$ ($\phi$) corresponds to a radial (azimuthal) linear optical dipole. Inset: magnified view of $F_{\mathrm{p}}$ for the structure without DBRs.}
\label{fig:purcells}
\end{figure}

The predicted Purcell factor for both radial and azimuthal dipoles as a function of the offset displacement $\rho$ is presented in Fig.~\ref{fig:purcells}. The structure with no DBRs exhibits no field enhancement at all (F$_{\mathrm{p}} < 1$), while the hourglass including DBRs features a maximum value of $F_{\mathrm{p}} \sim 45$ for $\rho = 0$. For both structures and dipole orientations, the enhancement drops with $\rho$, essentially following the electric field profile of the lateral profile of the fundamental HE$_{11}$ mode. 

\begin{figure}[hbtp] 
  \includegraphics[width=\linewidth]{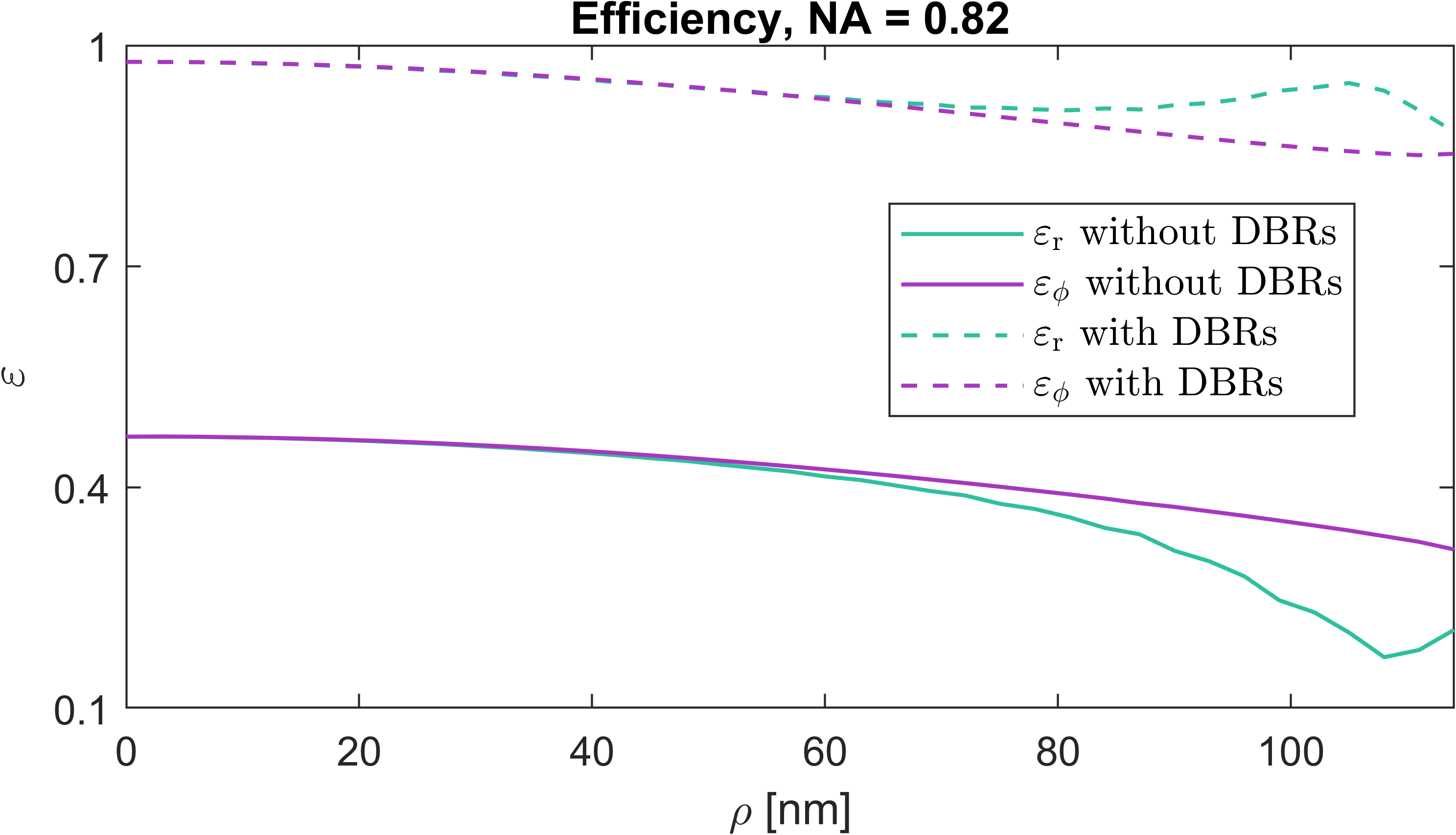}
\caption{Computed collection efficiency $\varepsilon$ for a radial (azimuthal) linear optical dipole as a function of its radial offset position $\rho$. The solid curves give the extraction efficiency accounting for a Gaussian overlap.}
\label{fig:effs}
\end{figure}

However, thanks to suppression \cite{gaal_near-unity_2022,wang_suppression_2021} of the background emission occurring for the small diameter $2R_{0} = 228\ \mathrm{nm}$ at the position of the QD, we have $\Gamma_{\mathrm{B}} \sim 0.05 \Gamma_{\mathrm{Bulk}}$ for a QD in the center, and the hourglass thus does not rely on strong Purcell enhancement to maintain a high efficiency. Indeed, the computed efficiency presented in Fig.~\ref{fig:effs} remains large ($\varepsilon > 0.85$) for the hourglass with DBRs even for large displacements $\rho$, where the Purcell factor is modest, indicating good tolerance towards spatial misalignment. Similar tolerance is observed in the absence of DBRs, albeit for corresponding lower efficiency. The different behaviors for the two dipole orientations in Fig.~\ref{fig:effs} for large $\rho$ result from the different boundary conditions experienced by the two electric field components at the semiconductor-air interface. 

\section{Results}

\begin{figure}[hb]
\centering
\includegraphics[width=\linewidth]{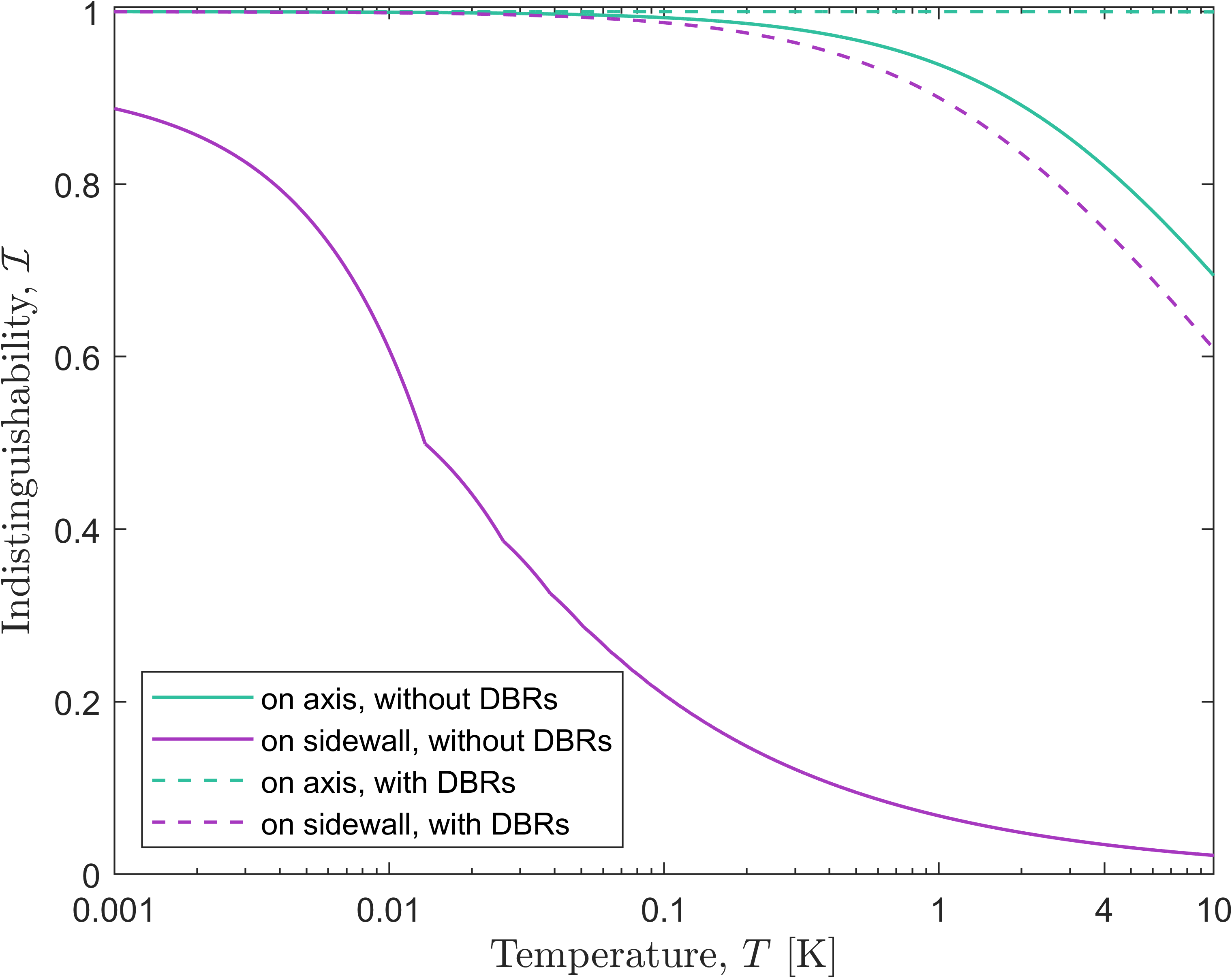}
\caption{Photon indistinguishability $\mathcal{I}$ with respect to the operation temperature for a QD located either perfectly on the neutral axis or on the sidewall of the structure for the two contemplated designs (with and without DBRs).}
\label{fig:indis}
\end{figure}

We now present the photon indistinguishability predicted from our model taking into account the discrete set of mechanical modes as a function of temperature in Fig.~\ref{fig:indis}. In the absence of DBRs, the indistinguishability is indeed strongly decreased due to interaction with flexural and longitudinal vibration modes. For a QD approaching the sidewall, it drops below 0.1 at \SI{4}{\kelvin} in agreement with previous results \cite{artioli_design_2019,tighineanu_phonon_2018}. However for the full hourglass structure including DBRs, we observe that $\mathcal{I} > 0.99$ over the entire temperature range for a QD in the center, indicating excellent resilience of the hourglass structure towards phonon decoherence in spite of the low-dimensional nature of the geometry. We attribute this surprisingly good performance to the significant Purcell enhancement offered by the cavity. Indeed, the indistinguishability is reduced for a QD on the sidewall due to the low $F_{\mathrm{p}}$ for increasing $\rho$ as discussed above. 

We summarize our predicted SPS figures of merit for the four configurations, on axis/sidewall and with/without DBRs, at \SI{4}{\kelvin} in Table~\ref{table:table1}. The lack of DBRs and Purcell enhancement results in reduced efficiency with some tolerance towards spatial misalignment. 
The flexural mechanical modes of the structure,  representing the primary source of decoherence, do not couple with a QD in the center, resulting in indistinguishability of 0.821 at \SI{4}{\kelvin} comparable to that in a bulk medium \cite{iles-smith_phonon_2017}. For a displaced QD at the sidewall, the interaction with these flexural modes almost completely ruins the indistinguishability for the structure without DBRs. In contrast, for the full hourglass structure with DBRs, this picture completely changes, and near-unity performance is obtained on-axis with $\varepsilon = 0.978$ and $\mathcal{I} = 0.9999$. 

\begin{table}[ht]
\centering
\begin{tabular}{ccccc}
\hline \hline
\addlinespace[0.25em]
& \multicolumn{2}{c}{\textit{Without DBRs}} & \multicolumn{2}{c}{\textit{With DBRs}} \\ \hline
\multicolumn{1}{c}{Figure of merit} & On axis & On sidewall & On axis & On sidewall \\ \addlinespace[0.125em] \hline
$F_{\mathrm{p}}$ & 0.91 & 0.08 & 44.20 & 3.66 \\
$\varepsilon$ & 0.469 & 0.207 & 0.978  & 0.883 \\ 
$\mathcal{I}$ & 0.821 & 0.034 & 0.9999 & 0.748 \\
$\varepsilon\mathcal{I}$ & 0.385 & 0.007 & \textbf{0.978} & 0.660 \\ \hline
Configuration & (a) & (b) & (c) & (d) \\ \hline \hline
\end{tabular}
\caption{Overview of SPS figures of merit for a radial dipole at $T$ = \SI{4}{\kelvin} with $\mathrm{NA}=0.82$. Boldface denotes the product of efficiency and indistinguishability for the full hourglass structure with DBRs, which reaches a maximum value of $\varepsilon\mathcal{I} = 0.978$.} 
\label{table:table1}
\end{table}

While the tolerance of the indistinguishability towards spatial misalignment is not as good as for the efficiency, we still observe that a product $\mathcal{I} \varepsilon = 0.66$ is obtained for a QD on the sidewall. Clearly, in practice, other effects, such as the detrimental influence of defect states near the semiconductor-air surface, will reduce this performance. Nevertheless, we observe that in the presence of cQED effects, the high aspect ratio and the 1D mechanical nature of the hourglass structure do not necessarily lead to strong phonon-induced decoherence, and the conclusion that background emission engineering \cite{gaal_near-unity_2022} can circumvent the trade-off \cite{iles-smith_phonon_2017} between efficiency and indistinguishability remains valid even for a mechanical structure of low dimensionality.

\section{Conclusion}

To conclude, we have implemented a model for phonon-induced decoherence fully taking into account the discrete set of mechanical modes of a low-dimensional photonic structure, and we have used it to evaluate the achievable indistinguishability for the hourglass SPS. While the indistinguishability of the bare structure without DBRs is indeed strongly affected by, in particular, the set of flexural vibration modes in consistency with the literature, our model predicts a surprisingly high indistinguishability $\mathcal{I}=0.9999$ for the full hourglass structure for an on-axis QD. We attribute this high indistinguishability not to the mechanical effects but rather to the strong Purcell enhancement introduced by the cavity. Furthermore, we show that not only the efficiency but also the indistinguishability displays reasonable tolerance towards spatial misalignment of the QD. Thus, our work confirms that the performance of a low-dimensional structure, here the hourglass SPS, can indeed match or outperform more conventional SPS designs.

\section*{Supplementary Material}
This supplementary material covers the investigation of thermally induced vibrations, where we present the material properties utilized in the FEM simulations and the extracted parameters of the vibration modes for the photonic hourglass structure, considering both configurations: with and without DBRs.

\section*{Acknowledgements}
The authors acknowledge support from the European Union’s Horizon 2020 Research and Innovation Programme under the Marie Skłodowska-Curie Grant (Agreement No. 861097) and from the European Research Council (ERC-CoG ``Unity", Grant No. 865230).

\section*{Author Declarations}
\subsection*{Conflict of Interest}
The authors have no conflicts to disclose.

\subsection*{Author Contributions}
\textbf{José Ferreira Neto:} Data curation (lead), Formal analysis (equal), Investigation (lead), Methodology (equal), Software (equal), Visualization (equal), Writing - original draft (lead), Writing - review \& editing (equal). \textbf{Benedek Gaál:} Investigation (equal), Software (equal), Validation (equal), Writing - review \& editing (supporting). \textbf{Luca Vannucci:} Conceptualization (equal), Methodology (supporting), Project administration (supporting), Supervision (equal), Writing - original draft (supporting), Writing - review \& editing (equal). \textbf{Niels Gregersen:} Conceptualization (lead), Funding acquisition (lead), Methodology (lead), Project administration (lead), Resources (lead), Supervision (lead), Writing - original draft (supporting), Writing - review \& editing (lead).

\section*{Data availability}
The data that support the findings of this study are available
from the corresponding authors upon reasonable request.

\appendix
\twocolumngrid
\section{Independent boson model}
\label{section:appendix_A}

The emission spectrum $S(\omega)$ is given by \cite{artioli_design_2019}

\begin{equation}
\label{eqn:1}
S(\omega)=\mathrm{Re}\int_{0}^{\infty}\mathrm{d}t\ P(t)\ e^{-\Gamma t/2} e^{-i\omega t},
\end{equation}
where $P(t)$ denotes the coherence evolution of the system, which in turn, can be expressed as a function of the phonon propagator $\Phi(t)$

\begin{equation}
\label{eqn:2}
\begin{aligned}
P(t) &= e^{\Phi(t)} \\
&= \prod_{m} P_{m}(t) \\
&= \prod_{m} \exp[-\theta_{m}^{2}\sin^{2}(\omega_{m}t/2)-\eta_{m}^{2}(1-e^{-i\omega_{m}t})].
\end{aligned}
\end{equation}

Upon inspection of Eq.~(\ref{eqn:2}), we can verify that the phonon propagator $\Phi(t)$ corresponds to

\begin{equation}
\label{eqn:3}
\Phi(t)=\sum_{m}-\theta_{m}^{2}\sin^{2}(\omega_{m}t/2)-\eta_{m}^{2}(1-e^{-i\omega_{m}t}).
\end{equation}
\\
Along with the phonon occupation number $N_{m}$ given by Bose-Einstein statistics, we will employ the following definitions for the coupling parameters $\theta_{m}^{2}$ and $\eta_{m}^{2}$

\begin{empheq}[left=\empheqlbrace]{align}
&\theta_{m}^{2}=4\eta_{m}^{2}N_{m} \label{eqn:4-1}, \\
&\eta_{m}^{2}=(g_{m}/\omega_{m})^{2} \label{eqn:4-2}, \\
&N_{m}=[\exp(\hbar\omega_{m}/k_{\mathrm{B}}T)-1]^{-1}. \label{eqn:4-3}
\end{empheq}

Here, $k_{\mathrm{B}}$ corresponds to the Boltzmann constant and $T$ is the bath temperature. Also, we resort to the given trigonometric relations

\begin{empheq}[left=\empheqlbrace]{align}
&\sin^{2}(\omega_{m}t/2)=[1-\cos(\omega_{m}t)]/2, \label{eqn:5-1} \\
&\coth(\hbar\omega_{m}/2k_{\mathrm{B}}T)=1+2N_{m}. \label{eqn:5-2}
\end{empheq}
\\
Now, making use of Eqs.~(\ref{eqn:4-1}), (\ref{eqn:4-2}), and (\ref{eqn:5-1}) in Eq.~(\ref{eqn:3})

\begin{equation}
\label{eqn:6}
\Phi(t)=\sum_{m}\frac{g_{m}^{2}}{\omega_{m}^{2}}\biggl\{-2N_{m}[1-\cos(\omega_{m}t)]-1+e^{-i\omega_{m}t}\biggr\}.
\end{equation}

Following, by isolating $N_{m}$ from (\ref{eqn:5-2}) and plugging it into (\ref{eqn:6}), we are left with

\begin{equation}
\label{eqn:7}
\begin{aligned}
\Phi(t) &=\sum_{m}\frac{g_{m}^{2}}{\omega_{m}^{2}}\biggl\{-[\coth(\hbar\omega_{m}/2k_{\mathrm{B}}T)-1][1-\cos(\omega_{m}t)] \biggr.\\
&\quad\biggl.-1+e^{-i\omega_{m}t}\biggr\}.
\end{aligned}
\end{equation}

Lastly, developing the expression above, we can bring it to the following final form

\begin{equation}
\label{eqn:8}
\begin{aligned}
\Phi(t) &=\sum_{m}\frac{g_{m}^{2}}{\omega_{m}^{2}}\biggl\{\coth\left(\frac{\hbar\omega_{m}}{2k_{\mathrm{B}}T}\right)[\cos(\omega_{m}t)-1] \biggr.\\
&\quad\biggl.-i\sin(\omega_{m}t)\biggr\},
\end{aligned}
\end{equation}
which corresponds to the standard phonon propagator expression derived from the independent boson model \cite{nazir_modelling_2016}.

\bibliography{main}

\begin{thebibliography}{35}%
\makeatletter
\providecommand \@ifxundefined [1]{%
 \@ifx{#1\undefined}
}%
\providecommand \@ifnum [1]{%
 \ifnum #1\expandafter \@firstoftwo
 \else \expandafter \@secondoftwo
 \fi
}%
\providecommand \@ifx [1]{%
 \ifx #1\expandafter \@firstoftwo
 \else \expandafter \@secondoftwo
 \fi
}%
\providecommand \natexlab [1]{#1}%
\providecommand \enquote  [1]{``#1''}%
\providecommand \bibnamefont  [1]{#1}%
\providecommand \bibfnamefont [1]{#1}%
\providecommand \citenamefont [1]{#1}%
\providecommand \href@noop [0]{\@secondoftwo}%
\providecommand \href [0]{\begingroup \@sanitize@url \@href}%
\providecommand \@href[1]{\@@startlink{#1}\@@href}%
\providecommand \@@href[1]{\endgroup#1\@@endlink}%
\providecommand \@sanitize@url [0]{\catcode `\\12\catcode `\$12\catcode `\&12\catcode `\#12\catcode `\^12\catcode `\_12\catcode `\%12\relax}%
\providecommand \@@startlink[1]{}%
\providecommand \@@endlink[0]{}%
\providecommand \url  [0]{\begingroup\@sanitize@url \@url }%
\providecommand \@url [1]{\endgroup\@href {#1}{\urlprefix }}%
\providecommand \urlprefix  [0]{URL }%
\providecommand \Eprint [0]{\href }%
\providecommand \doibase [0]{https://doi.org/}%
\providecommand \selectlanguage [0]{\@gobble}%
\providecommand \bibinfo  [0]{\@secondoftwo}%
\providecommand \bibfield  [0]{\@secondoftwo}%
\providecommand \translation [1]{[#1]}%
\providecommand \BibitemOpen [0]{}%
\providecommand \bibitemStop [0]{}%
\providecommand \bibitemNoStop [0]{.\EOS\space}%
\providecommand \EOS [0]{\spacefactor3000\relax}%
\providecommand \BibitemShut  [1]{\csname bibitem#1\endcsname}%
\let\auto@bib@innerbib\@empty
\bibitem [{\citenamefont {Feynman}(1982)}]{feynman_simulating_1982}%
  \BibitemOpen
  \bibfield  {author} {\bibinfo {author} {\bibfnamefont {R.~P.}\ \bibnamefont {Feynman}},\ }\bibfield  {title} {{\selectlanguage {en}\bibinfo {title} {Simulating physics with computers}},\ }\href {https://doi.org/10.1007/BF02650179} {\bibfield  {journal} {\bibinfo  {journal} {Int J Theor Phys}\ }\textbf {\bibinfo {volume} {21}},\ \bibinfo {pages} {467} (\bibinfo {year} {1982})}\BibitemShut {NoStop}%
\bibitem [{noa(2022)}]{noauthor_40_2022}%
  \BibitemOpen
  \bibfield  {title} {{\selectlanguage {en}\bibinfo {title} {40 years of quantum computing}},\ }\href {https://doi.org/10.1038/s42254-021-00410-6} {\bibfield  {journal} {\bibinfo  {journal} {Nat Rev Phys}\ }\textbf {\bibinfo {volume} {4}},\ \bibinfo {pages} {1} (\bibinfo {year} {2022})}\BibitemShut {NoStop}%
\bibitem [{\citenamefont {Politi}\ \emph {et~al.}(2009)\citenamefont {Politi}, \citenamefont {Matthews},\ and\ \citenamefont {O'Brien}}]{politi_shors_2009}%
  \BibitemOpen
  \bibfield  {author} {\bibinfo {author} {\bibfnamefont {A.}~\bibnamefont {Politi}}, \bibinfo {author} {\bibfnamefont {J.~C.~F.}\ \bibnamefont {Matthews}},\ and\ \bibinfo {author} {\bibfnamefont {J.~L.}\ \bibnamefont {O'Brien}},\ }\bibfield  {title} {{\selectlanguage {en}\bibinfo {title} {Shor’s {Quantum} {Factoring} {Algorithm} on a {Photonic} {Chip}}},\ }\href {https://doi.org/10.1126/science.1173731} {\bibfield  {journal} {\bibinfo  {journal} {Science}\ }\textbf {\bibinfo {volume} {325}},\ \bibinfo {pages} {1221} (\bibinfo {year} {2009})}\BibitemShut {NoStop}%
\bibitem [{\citenamefont {Crespi}\ \emph {et~al.}(2011)\citenamefont {Crespi}, \citenamefont {Ramponi}, \citenamefont {Osellame}, \citenamefont {Sansoni}, \citenamefont {Bongioanni}, \citenamefont {Sciarrino}, \citenamefont {Vallone},\ and\ \citenamefont {Mataloni}}]{crespi_integrated_2011}%
  \BibitemOpen
  \bibfield  {author} {\bibinfo {author} {\bibfnamefont {A.}~\bibnamefont {Crespi}}, \bibinfo {author} {\bibfnamefont {R.}~\bibnamefont {Ramponi}}, \bibinfo {author} {\bibfnamefont {R.}~\bibnamefont {Osellame}}, \bibinfo {author} {\bibfnamefont {L.}~\bibnamefont {Sansoni}}, \bibinfo {author} {\bibfnamefont {I.}~\bibnamefont {Bongioanni}}, \bibinfo {author} {\bibfnamefont {F.}~\bibnamefont {Sciarrino}}, \bibinfo {author} {\bibfnamefont {G.}~\bibnamefont {Vallone}},\ and\ \bibinfo {author} {\bibfnamefont {P.}~\bibnamefont {Mataloni}},\ }\bibfield  {title} {{\selectlanguage {en}\bibinfo {title} {Integrated photonic quantum gates for polarization qubits}},\ }\href {https://doi.org/10.1038/ncomms1570} {\bibfield  {journal} {\bibinfo  {journal} {Nat Commun}\ }\textbf {\bibinfo {volume} {2}},\ \bibinfo {pages} {566} (\bibinfo {year} {2011})}\BibitemShut {NoStop}%
\bibitem [{\citenamefont {Humphreys}\ \emph {et~al.}(2013)\citenamefont {Humphreys}, \citenamefont {Metcalf}, \citenamefont {Spring}, \citenamefont {Moore}, \citenamefont {Jin}, \citenamefont {Barbieri}, \citenamefont {Kolthammer},\ and\ \citenamefont {Walmsley}}]{humphreys_linear_2013}%
  \BibitemOpen
  \bibfield  {author} {\bibinfo {author} {\bibfnamefont {P.~C.}\ \bibnamefont {Humphreys}}, \bibinfo {author} {\bibfnamefont {B.~J.}\ \bibnamefont {Metcalf}}, \bibinfo {author} {\bibfnamefont {J.~B.}\ \bibnamefont {Spring}}, \bibinfo {author} {\bibfnamefont {M.}~\bibnamefont {Moore}}, \bibinfo {author} {\bibfnamefont {X.-M.}\ \bibnamefont {Jin}}, \bibinfo {author} {\bibfnamefont {M.}~\bibnamefont {Barbieri}}, \bibinfo {author} {\bibfnamefont {W.~S.}\ \bibnamefont {Kolthammer}},\ and\ \bibinfo {author} {\bibfnamefont {I.~A.}\ \bibnamefont {Walmsley}},\ }\bibfield  {title} {{\selectlanguage {en}\bibinfo {title} {Linear {Optical} {Quantum} {Computing} in a {Single} {Spatial} {Mode}}},\ }\href {https://doi.org/10.1103/PhysRevLett.111.150501} {\bibfield  {journal} {\bibinfo  {journal} {Phys. Rev. Lett.}\ }\textbf {\bibinfo {volume} {111}},\ \bibinfo {pages} {150501} (\bibinfo {year} {2013})}\BibitemShut {NoStop}%
\bibitem [{\citenamefont {Zhong}\ \emph {et~al.}(2020)\citenamefont {Zhong}, \citenamefont {Wang}, \citenamefont {Deng}, \citenamefont {Chen}, \citenamefont {Peng}, \citenamefont {Luo}, \citenamefont {Qin}, \citenamefont {Wu}, \citenamefont {Ding}, \citenamefont {Hu}, \citenamefont {Hu}, \citenamefont {Yang}, \citenamefont {Zhang}, \citenamefont {Li}, \citenamefont {Li}, \citenamefont {Jiang}, \citenamefont {Gan}, \citenamefont {Yang}, \citenamefont {You}, \citenamefont {Wang}, \citenamefont {Li}, \citenamefont {Liu}, \citenamefont {Lu},\ and\ \citenamefont {Pan}}]{zhong_quantum_2020}%
  \BibitemOpen
  \bibfield  {author} {\bibinfo {author} {\bibfnamefont {H.-S.}\ \bibnamefont {Zhong}}, \bibinfo {author} {\bibfnamefont {H.}~\bibnamefont {Wang}}, \bibinfo {author} {\bibfnamefont {Y.-H.}\ \bibnamefont {Deng}}, \bibinfo {author} {\bibfnamefont {M.-C.}\ \bibnamefont {Chen}}, \bibinfo {author} {\bibfnamefont {L.-C.}\ \bibnamefont {Peng}}, \bibinfo {author} {\bibfnamefont {Y.-H.}\ \bibnamefont {Luo}}, \bibinfo {author} {\bibfnamefont {J.}~\bibnamefont {Qin}}, \bibinfo {author} {\bibfnamefont {D.}~\bibnamefont {Wu}}, \bibinfo {author} {\bibfnamefont {X.}~\bibnamefont {Ding}}, \bibinfo {author} {\bibfnamefont {Y.}~\bibnamefont {Hu}}, \bibinfo {author} {\bibfnamefont {P.}~\bibnamefont {Hu}}, \bibinfo {author} {\bibfnamefont {X.-Y.}\ \bibnamefont {Yang}}, \bibinfo {author} {\bibfnamefont {W.-J.}\ \bibnamefont {Zhang}}, \bibinfo {author} {\bibfnamefont {H.}~\bibnamefont {Li}}, \bibinfo {author} {\bibfnamefont {Y.}~\bibnamefont {Li}}, \bibinfo {author} {\bibfnamefont {X.}~\bibnamefont {Jiang}}, \bibinfo {author}
  {\bibfnamefont {L.}~\bibnamefont {Gan}}, \bibinfo {author} {\bibfnamefont {G.}~\bibnamefont {Yang}}, \bibinfo {author} {\bibfnamefont {L.}~\bibnamefont {You}}, \bibinfo {author} {\bibfnamefont {Z.}~\bibnamefont {Wang}}, \bibinfo {author} {\bibfnamefont {L.}~\bibnamefont {Li}}, \bibinfo {author} {\bibfnamefont {N.-L.}\ \bibnamefont {Liu}}, \bibinfo {author} {\bibfnamefont {C.-Y.}\ \bibnamefont {Lu}},\ and\ \bibinfo {author} {\bibfnamefont {J.-W.}\ \bibnamefont {Pan}},\ }\bibfield  {title} {{\selectlanguage {en}\bibinfo {title} {Quantum computational advantage using photons}},\ }\href {https://doi.org/10.1126/science.abe8770} {\bibfield  {journal} {\bibinfo  {journal} {Science}\ }\textbf {\bibinfo {volume} {370}},\ \bibinfo {pages} {1460} (\bibinfo {year} {2020})}\BibitemShut {NoStop}%
\bibitem [{\citenamefont {Gregersen}\ \emph {et~al.}(2017)\citenamefont {Gregersen}, \citenamefont {McCutcheon},\ and\ \citenamefont {Mørk}}]{piprek_handbook_2017}%
  \BibitemOpen
  \bibfield  {author} {\bibinfo {author} {\bibfnamefont {N.}~\bibnamefont {Gregersen}}, \bibinfo {author} {\bibfnamefont {D.~P.~S.}\ \bibnamefont {McCutcheon}},\ and\ \bibinfo {author} {\bibfnamefont {J.}~\bibnamefont {Mørk}},\ }\href {https://doi.org/10.4324/9781315152318} {{\selectlanguage {en}\emph {\bibinfo {title} {Single-photon sources in Handbook of Optoelectronic Device Modeling and Simulation: Lasers, Modulators, Photodetectors, Solar Cells, and Numerical Methods}}}},\ \bibinfo {edition} {1st}\ ed.,\ edited by\ \bibinfo {editor} {\bibfnamefont {J.}~\bibnamefont {Piprek}}\ (\bibinfo  {publisher} {CRC Press},\ \bibinfo {address} {Boca Raton, FL : CRC Press, Taylor \& Francis Group},\ \bibinfo {year} {2017})\ \bibinfo {note} {vol. 2, Chap. 46, pp. 585–607}\BibitemShut {NoStop}%
\bibitem [{\citenamefont {Heindel}\ \emph {et~al.}(2023)\citenamefont {Heindel}, \citenamefont {Kim}, \citenamefont {Gregersen}, \citenamefont {Rastelli},\ and\ \citenamefont {Reitzenstein}}]{heindel_quantum_2023}%
  \BibitemOpen
  \bibfield  {author} {\bibinfo {author} {\bibfnamefont {T.}~\bibnamefont {Heindel}}, \bibinfo {author} {\bibfnamefont {J.-H.}\ \bibnamefont {Kim}}, \bibinfo {author} {\bibfnamefont {N.}~\bibnamefont {Gregersen}}, \bibinfo {author} {\bibfnamefont {A.}~\bibnamefont {Rastelli}},\ and\ \bibinfo {author} {\bibfnamefont {S.}~\bibnamefont {Reitzenstein}},\ }\bibfield  {title} {{\selectlanguage {en}\bibinfo {title} {Quantum dots for photonic quantum information technology}},\ }\href {https://doi.org/10.1364/AOP.490091} {\bibfield  {journal} {\bibinfo  {journal} {Adv. Opt. Photon.}\ }\textbf {\bibinfo {volume} {15}},\ \bibinfo {pages} {613} (\bibinfo {year} {2023})}\BibitemShut {NoStop}%
\bibitem [{\citenamefont {Hong}\ \emph {et~al.}(1987)\citenamefont {Hong}, \citenamefont {Ou},\ and\ \citenamefont {Mandel}}]{hong_measurement_1987}%
  \BibitemOpen
  \bibfield  {author} {\bibinfo {author} {\bibfnamefont {C.~K.}\ \bibnamefont {Hong}}, \bibinfo {author} {\bibfnamefont {Z.~Y.}\ \bibnamefont {Ou}},\ and\ \bibinfo {author} {\bibfnamefont {L.}~\bibnamefont {Mandel}},\ }\bibfield  {title} {{\selectlanguage {en}\bibinfo {title} {Measurement of subpicosecond time intervals between two photons by interference}},\ }\href {https://doi.org/10.1103/PhysRevLett.59.2044} {\bibfield  {journal} {\bibinfo  {journal} {Phys. Rev. Lett.}\ }\textbf {\bibinfo {volume} {59}},\ \bibinfo {pages} {2044} (\bibinfo {year} {1987})}\BibitemShut {NoStop}%
\bibitem [{\citenamefont {Kwiat}\ \emph {et~al.}(1995)\citenamefont {Kwiat}, \citenamefont {Mattle}, \citenamefont {Weinfurter}, \citenamefont {Zeilinger}, \citenamefont {Sergienko},\ and\ \citenamefont {Shih}}]{kwiat_new_1995}%
  \BibitemOpen
  \bibfield  {author} {\bibinfo {author} {\bibfnamefont {P.~G.}\ \bibnamefont {Kwiat}}, \bibinfo {author} {\bibfnamefont {K.}~\bibnamefont {Mattle}}, \bibinfo {author} {\bibfnamefont {H.}~\bibnamefont {Weinfurter}}, \bibinfo {author} {\bibfnamefont {A.}~\bibnamefont {Zeilinger}}, \bibinfo {author} {\bibfnamefont {A.~V.}\ \bibnamefont {Sergienko}},\ and\ \bibinfo {author} {\bibfnamefont {Y.}~\bibnamefont {Shih}},\ }\bibfield  {title} {{\selectlanguage {en}\bibinfo {title} {New {High}-{Intensity} {Source} of {Polarization}-{Entangled} {Photon} {Pairs}}},\ }\href {https://doi.org/10.1103/PhysRevLett.75.4337} {\bibfield  {journal} {\bibinfo  {journal} {Phys. Rev. Lett.}\ }\textbf {\bibinfo {volume} {75}},\ \bibinfo {pages} {4337} (\bibinfo {year} {1995})}\BibitemShut {NoStop}%
\bibitem [{\citenamefont {Senellart}\ \emph {et~al.}(2017)\citenamefont {Senellart}, \citenamefont {Solomon},\ and\ \citenamefont {White}}]{senellart_high-performance_2017}%
  \BibitemOpen
  \bibfield  {author} {\bibinfo {author} {\bibfnamefont {P.}~\bibnamefont {Senellart}}, \bibinfo {author} {\bibfnamefont {G.}~\bibnamefont {Solomon}},\ and\ \bibinfo {author} {\bibfnamefont {A.}~\bibnamefont {White}},\ }\bibfield  {title} {{\selectlanguage {en}\bibinfo {title} {High-performance semiconductor quantum-dot single-photon sources}},\ }\href {https://doi.org/10.1038/nnano.2017.218} {\bibfield  {journal} {\bibinfo  {journal} {Nature Nanotech}\ }\textbf {\bibinfo {volume} {12}},\ \bibinfo {pages} {1026} (\bibinfo {year} {2017})}\BibitemShut {NoStop}%
\bibitem [{\citenamefont {Wang}\ \emph {et~al.}(2019)\citenamefont {Wang}, \citenamefont {He}, \citenamefont {Chung}, \citenamefont {Hu}, \citenamefont {Yu}, \citenamefont {Chen}, \citenamefont {Ding}, \citenamefont {Chen}, \citenamefont {Qin}, \citenamefont {Yang}, \citenamefont {Liu}, \citenamefont {Duan}, \citenamefont {Li}, \citenamefont {Gerhardt}, \citenamefont {Winkler}, \citenamefont {Jurkat}, \citenamefont {Wang}, \citenamefont {Gregersen}, \citenamefont {Huo}, \citenamefont {Dai}, \citenamefont {Yu}, \citenamefont {Höfling}, \citenamefont {Lu},\ and\ \citenamefont {Pan}}]{wang_towards_2019}%
  \BibitemOpen
  \bibfield  {author} {\bibinfo {author} {\bibfnamefont {H.}~\bibnamefont {Wang}}, \bibinfo {author} {\bibfnamefont {Y.-M.}\ \bibnamefont {He}}, \bibinfo {author} {\bibfnamefont {T.-H.}\ \bibnamefont {Chung}}, \bibinfo {author} {\bibfnamefont {H.}~\bibnamefont {Hu}}, \bibinfo {author} {\bibfnamefont {Y.}~\bibnamefont {Yu}}, \bibinfo {author} {\bibfnamefont {S.}~\bibnamefont {Chen}}, \bibinfo {author} {\bibfnamefont {X.}~\bibnamefont {Ding}}, \bibinfo {author} {\bibfnamefont {M.-C.}\ \bibnamefont {Chen}}, \bibinfo {author} {\bibfnamefont {J.}~\bibnamefont {Qin}}, \bibinfo {author} {\bibfnamefont {X.}~\bibnamefont {Yang}}, \bibinfo {author} {\bibfnamefont {R.-Z.}\ \bibnamefont {Liu}}, \bibinfo {author} {\bibfnamefont {Z.-C.}\ \bibnamefont {Duan}}, \bibinfo {author} {\bibfnamefont {J.-P.}\ \bibnamefont {Li}}, \bibinfo {author} {\bibfnamefont {S.}~\bibnamefont {Gerhardt}}, \bibinfo {author} {\bibfnamefont {K.}~\bibnamefont {Winkler}}, \bibinfo {author} {\bibfnamefont {J.}~\bibnamefont {Jurkat}}, \bibinfo {author}
  {\bibfnamefont {L.-J.}\ \bibnamefont {Wang}}, \bibinfo {author} {\bibfnamefont {N.}~\bibnamefont {Gregersen}}, \bibinfo {author} {\bibfnamefont {Y.-H.}\ \bibnamefont {Huo}}, \bibinfo {author} {\bibfnamefont {Q.}~\bibnamefont {Dai}}, \bibinfo {author} {\bibfnamefont {S.}~\bibnamefont {Yu}}, \bibinfo {author} {\bibfnamefont {S.}~\bibnamefont {Höfling}}, \bibinfo {author} {\bibfnamefont {C.-Y.}\ \bibnamefont {Lu}},\ and\ \bibinfo {author} {\bibfnamefont {J.-W.}\ \bibnamefont {Pan}},\ }\bibfield  {title} {{\selectlanguage {en}\bibinfo {title} {Towards optimal single-photon sources from polarized microcavities}},\ }\href {https://doi.org/10.1038/s41566-019-0494-3} {\bibfield  {journal} {\bibinfo  {journal} {Nat. Photonics}\ }\textbf {\bibinfo {volume} {13}},\ \bibinfo {pages} {770} (\bibinfo {year} {2019})}\BibitemShut {NoStop}%
\bibitem [{\citenamefont {Tomm}\ \emph {et~al.}(2021)\citenamefont {Tomm}, \citenamefont {Javadi}, \citenamefont {Antoniadis}, \citenamefont {Najer}, \citenamefont {Löbl}, \citenamefont {Korsch}, \citenamefont {Schott}, \citenamefont {Valentin}, \citenamefont {Wieck}, \citenamefont {Ludwig},\ and\ \citenamefont {Warburton}}]{tomm_bright_2021}%
  \BibitemOpen
  \bibfield  {author} {\bibinfo {author} {\bibfnamefont {N.}~\bibnamefont {Tomm}}, \bibinfo {author} {\bibfnamefont {A.}~\bibnamefont {Javadi}}, \bibinfo {author} {\bibfnamefont {N.~O.}\ \bibnamefont {Antoniadis}}, \bibinfo {author} {\bibfnamefont {D.}~\bibnamefont {Najer}}, \bibinfo {author} {\bibfnamefont {M.~C.}\ \bibnamefont {Löbl}}, \bibinfo {author} {\bibfnamefont {A.~R.}\ \bibnamefont {Korsch}}, \bibinfo {author} {\bibfnamefont {R.}~\bibnamefont {Schott}}, \bibinfo {author} {\bibfnamefont {S.~R.}\ \bibnamefont {Valentin}}, \bibinfo {author} {\bibfnamefont {A.~D.}\ \bibnamefont {Wieck}}, \bibinfo {author} {\bibfnamefont {A.}~\bibnamefont {Ludwig}},\ and\ \bibinfo {author} {\bibfnamefont {R.~J.}\ \bibnamefont {Warburton}},\ }\bibfield  {title} {{\selectlanguage {en}\bibinfo {title} {A bright and fast source of coherent single photons}},\ }\href {https://doi.org/10.1038/s41565-020-00831-x} {\bibfield  {journal} {\bibinfo  {journal} {Nat. Nanotechnol.}\ }\textbf {\bibinfo {volume} {16}},\ \bibinfo {pages}
  {399} (\bibinfo {year} {2021})}\BibitemShut {NoStop}%
\bibitem [{\citenamefont {Maring}\ \emph {et~al.}(2024)\citenamefont {Maring}, \citenamefont {Fyrillas}, \citenamefont {Pont}, \citenamefont {Ivanov}, \citenamefont {Stepanov}, \citenamefont {Margaria}, \citenamefont {Hease}, \citenamefont {Pishchagin}, \citenamefont {Lemaître}, \citenamefont {Sagnes}, \citenamefont {Au}, \citenamefont {Boissier}, \citenamefont {Bertasi}, \citenamefont {Baert}, \citenamefont {Valdivia}, \citenamefont {Billard}, \citenamefont {Acar}, \citenamefont {Brieussel}, \citenamefont {Mezher}, \citenamefont {Wein}, \citenamefont {Salavrakos}, \citenamefont {Sinnott}, \citenamefont {Fioretto}, \citenamefont {Emeriau}, \citenamefont {Belabas}, \citenamefont {Mansfield}, \citenamefont {Senellart}, \citenamefont {Senellart},\ and\ \citenamefont {Somaschi}}]{maring_versatile_2024}%
  \BibitemOpen
  \bibfield  {author} {\bibinfo {author} {\bibfnamefont {N.}~\bibnamefont {Maring}}, \bibinfo {author} {\bibfnamefont {A.}~\bibnamefont {Fyrillas}}, \bibinfo {author} {\bibfnamefont {M.}~\bibnamefont {Pont}}, \bibinfo {author} {\bibfnamefont {E.}~\bibnamefont {Ivanov}}, \bibinfo {author} {\bibfnamefont {P.}~\bibnamefont {Stepanov}}, \bibinfo {author} {\bibfnamefont {N.}~\bibnamefont {Margaria}}, \bibinfo {author} {\bibfnamefont {W.}~\bibnamefont {Hease}}, \bibinfo {author} {\bibfnamefont {A.}~\bibnamefont {Pishchagin}}, \bibinfo {author} {\bibfnamefont {A.}~\bibnamefont {Lemaître}}, \bibinfo {author} {\bibfnamefont {I.}~\bibnamefont {Sagnes}}, \bibinfo {author} {\bibfnamefont {T.~H.}\ \bibnamefont {Au}}, \bibinfo {author} {\bibfnamefont {S.}~\bibnamefont {Boissier}}, \bibinfo {author} {\bibfnamefont {E.}~\bibnamefont {Bertasi}}, \bibinfo {author} {\bibfnamefont {A.}~\bibnamefont {Baert}}, \bibinfo {author} {\bibfnamefont {M.}~\bibnamefont {Valdivia}}, \bibinfo {author} {\bibfnamefont {M.}~\bibnamefont
  {Billard}}, \bibinfo {author} {\bibfnamefont {O.}~\bibnamefont {Acar}}, \bibinfo {author} {\bibfnamefont {A.}~\bibnamefont {Brieussel}}, \bibinfo {author} {\bibfnamefont {R.}~\bibnamefont {Mezher}}, \bibinfo {author} {\bibfnamefont {S.~C.}\ \bibnamefont {Wein}}, \bibinfo {author} {\bibfnamefont {A.}~\bibnamefont {Salavrakos}}, \bibinfo {author} {\bibfnamefont {P.}~\bibnamefont {Sinnott}}, \bibinfo {author} {\bibfnamefont {D.~A.}\ \bibnamefont {Fioretto}}, \bibinfo {author} {\bibfnamefont {P.-E.}\ \bibnamefont {Emeriau}}, \bibinfo {author} {\bibfnamefont {N.}~\bibnamefont {Belabas}}, \bibinfo {author} {\bibfnamefont {S.}~\bibnamefont {Mansfield}}, \bibinfo {author} {\bibfnamefont {P.}~\bibnamefont {Senellart}}, \bibinfo {author} {\bibfnamefont {J.}~\bibnamefont {Senellart}},\ and\ \bibinfo {author} {\bibfnamefont {N.}~\bibnamefont {Somaschi}},\ }\bibfield  {title} {{\selectlanguage {en}\bibinfo {title} {A versatile single-photon-based quantum computing platform}},\ }\href
  {https://doi.org/10.1038/s41566-024-01403-4} {\bibfield  {journal} {\bibinfo  {journal} {Nat. Photon.}\ }\textbf {\bibinfo {volume} {18}},\ \bibinfo {pages} {603} (\bibinfo {year} {2024})}\BibitemShut {NoStop}%
\bibitem [{\citenamefont {Iles-Smith}\ \emph {et~al.}(2017)\citenamefont {Iles-Smith}, \citenamefont {McCutcheon}, \citenamefont {Nazir},\ and\ \citenamefont {Mørk}}]{iles-smith_phonon_2017}%
  \BibitemOpen
  \bibfield  {author} {\bibinfo {author} {\bibfnamefont {J.}~\bibnamefont {Iles-Smith}}, \bibinfo {author} {\bibfnamefont {D.~P.~S.}\ \bibnamefont {McCutcheon}}, \bibinfo {author} {\bibfnamefont {A.}~\bibnamefont {Nazir}},\ and\ \bibinfo {author} {\bibfnamefont {J.}~\bibnamefont {Mørk}},\ }\bibfield  {title} {{\selectlanguage {en}\bibinfo {title} {Phonon scattering inhibits simultaneous near-unity efficiency and indistinguishability in semiconductor single-photon sources}},\ }\href {https://doi.org/10.1038/nphoton.2017.101} {\bibfield  {journal} {\bibinfo  {journal} {Nature Photon}\ }\textbf {\bibinfo {volume} {11}},\ \bibinfo {pages} {521} (\bibinfo {year} {2017})}\BibitemShut {NoStop}%
\bibitem [{\citenamefont {Wang}\ \emph {et~al.}(2020)\citenamefont {Wang}, \citenamefont {Denning}, \citenamefont {Gür}, \citenamefont {Lu},\ and\ \citenamefont {Gregersen}}]{wang_micropillar_2020}%
  \BibitemOpen
  \bibfield  {author} {\bibinfo {author} {\bibfnamefont {B.-Y.}\ \bibnamefont {Wang}}, \bibinfo {author} {\bibfnamefont {E.~V.}\ \bibnamefont {Denning}}, \bibinfo {author} {\bibfnamefont {U.~M.}\ \bibnamefont {Gür}}, \bibinfo {author} {\bibfnamefont {C.-Y.}\ \bibnamefont {Lu}},\ and\ \bibinfo {author} {\bibfnamefont {N.}~\bibnamefont {Gregersen}},\ }\bibfield  {title} {{\selectlanguage {en}\bibinfo {title} {Micropillar single-photon source design for simultaneous near-unity efficiency and indistinguishability}},\ }\href {https://doi.org/10.1103/PhysRevB.102.125301} {\bibfield  {journal} {\bibinfo  {journal} {Phys. Rev. B}\ }\textbf {\bibinfo {volume} {102}},\ \bibinfo {pages} {125301} (\bibinfo {year} {2020})}\BibitemShut {NoStop}%
\bibitem [{\citenamefont {Osterkryger}\ \emph {et~al.}(2019)\citenamefont {Osterkryger}, \citenamefont {Claudon}, \citenamefont {Gérard},\ and\ \citenamefont {Gregersen}}]{osterkryger_photonic_2019}%
  \BibitemOpen
  \bibfield  {author} {\bibinfo {author} {\bibfnamefont {A.~D.}\ \bibnamefont {Osterkryger}}, \bibinfo {author} {\bibfnamefont {J.}~\bibnamefont {Claudon}}, \bibinfo {author} {\bibfnamefont {J.-M.}\ \bibnamefont {Gérard}},\ and\ \bibinfo {author} {\bibfnamefont {N.}~\bibnamefont {Gregersen}},\ }\bibfield  {title} {{\selectlanguage {en}\bibinfo {title} {Photonic “hourglass” design for efficient quantum light emission}},\ }\href {https://doi.org/10.1364/OL.44.002617} {\bibfield  {journal} {\bibinfo  {journal} {Opt. Lett.}\ }\textbf {\bibinfo {volume} {44}},\ \bibinfo {pages} {2617} (\bibinfo {year} {2019})}\BibitemShut {NoStop}%
\bibitem [{\citenamefont {Gaál}\ \emph {et~al.}(2022)\citenamefont {Gaál}, \citenamefont {Jacobsen}, \citenamefont {Vannucci}, \citenamefont {Claudon}, \citenamefont {Gérard},\ and\ \citenamefont {Gregersen}}]{gaal_near-unity_2022}%
  \BibitemOpen
  \bibfield  {author} {\bibinfo {author} {\bibfnamefont {B.}~\bibnamefont {Gaál}}, \bibinfo {author} {\bibfnamefont {M.~A.}\ \bibnamefont {Jacobsen}}, \bibinfo {author} {\bibfnamefont {L.}~\bibnamefont {Vannucci}}, \bibinfo {author} {\bibfnamefont {J.}~\bibnamefont {Claudon}}, \bibinfo {author} {\bibfnamefont {J.-M.}\ \bibnamefont {Gérard}},\ and\ \bibinfo {author} {\bibfnamefont {N.}~\bibnamefont {Gregersen}},\ }\bibfield  {title} {{\selectlanguage {en}\bibinfo {title} {Near-unity efficiency and photon indistinguishability for the “hourglass” single-photon source using suppression of the background emission}},\ }\href {https://doi.org/10.1063/5.0107624} {\bibfield  {journal} {\bibinfo  {journal} {Applied Physics Letters}\ }\textbf {\bibinfo {volume} {121}},\ \bibinfo {pages} {170501} (\bibinfo {year} {2022})}\BibitemShut {NoStop}%
\bibitem [{\citenamefont {Wang}\ \emph {et~al.}(2021)\citenamefont {Wang}, \citenamefont {Häyrynen}, \citenamefont {Vannucci}, \citenamefont {Jacobsen}, \citenamefont {Lu},\ and\ \citenamefont {Gregersen}}]{wang_suppression_2021}%
  \BibitemOpen
  \bibfield  {author} {\bibinfo {author} {\bibfnamefont {B.-Y.}\ \bibnamefont {Wang}}, \bibinfo {author} {\bibfnamefont {T.}~\bibnamefont {Häyrynen}}, \bibinfo {author} {\bibfnamefont {L.}~\bibnamefont {Vannucci}}, \bibinfo {author} {\bibfnamefont {M.~A.}\ \bibnamefont {Jacobsen}}, \bibinfo {author} {\bibfnamefont {C.-Y.}\ \bibnamefont {Lu}},\ and\ \bibinfo {author} {\bibfnamefont {N.}~\bibnamefont {Gregersen}},\ }\bibfield  {title} {{\selectlanguage {en}\bibinfo {title} {Suppression of background emission for efficient single-photon generation in micropillar cavities}},\ }\href {https://doi.org/10.1063/5.0044018} {\bibfield  {journal} {\bibinfo  {journal} {Applied Physics Letters}\ }\textbf {\bibinfo {volume} {118}},\ \bibinfo {pages} {114003} (\bibinfo {year} {2021})}\BibitemShut {NoStop}%
\bibitem [{\citenamefont {Tighineanu}\ \emph {et~al.}(2018)\citenamefont {Tighineanu}, \citenamefont {Dreeßen}, \citenamefont {Flindt}, \citenamefont {Lodahl},\ and\ \citenamefont {Sørensen}}]{tighineanu_phonon_2018}%
  \BibitemOpen
  \bibfield  {author} {\bibinfo {author} {\bibfnamefont {P.}~\bibnamefont {Tighineanu}}, \bibinfo {author} {\bibfnamefont {C.}~\bibnamefont {Dreeßen}}, \bibinfo {author} {\bibfnamefont {C.}~\bibnamefont {Flindt}}, \bibinfo {author} {\bibfnamefont {P.}~\bibnamefont {Lodahl}},\ and\ \bibinfo {author} {\bibfnamefont {A.}~\bibnamefont {Sørensen}},\ }\bibfield  {title} {{\selectlanguage {en}\bibinfo {title} {Phonon {Decoherence} of {Quantum} {Dots} in {Photonic} {Structures}: {Broadening} of the {Zero}-{Phonon} {Line} and the {Role} of {Dimensionality}}},\ }\href {https://doi.org/10.1103/PhysRevLett.120.257401} {\bibfield  {journal} {\bibinfo  {journal} {Phys. Rev. Lett.}\ }\textbf {\bibinfo {volume} {120}},\ \bibinfo {pages} {257401} (\bibinfo {year} {2018})}\BibitemShut {NoStop}%
\bibitem [{\citenamefont {Gregersen}\ \emph {et~al.}(2016)\citenamefont {Gregersen}, \citenamefont {McCutcheon}, \citenamefont {Mørk}, \citenamefont {Gérard},\ and\ \citenamefont {Claudon}}]{gregersen_broadband_2016}%
  \BibitemOpen
  \bibfield  {author} {\bibinfo {author} {\bibfnamefont {N.}~\bibnamefont {Gregersen}}, \bibinfo {author} {\bibfnamefont {D.~P.~S.}\ \bibnamefont {McCutcheon}}, \bibinfo {author} {\bibfnamefont {J.}~\bibnamefont {Mørk}}, \bibinfo {author} {\bibfnamefont {J.-M.}\ \bibnamefont {Gérard}},\ and\ \bibinfo {author} {\bibfnamefont {J.}~\bibnamefont {Claudon}},\ }\bibfield  {title} {{\selectlanguage {en}\bibinfo {title} {A broadband tapered nanocavity for efficient nonclassical light emission}},\ }\href {https://doi.org/10.1364/OE.24.020904} {\bibfield  {journal} {\bibinfo  {journal} {Opt. Express}\ }\textbf {\bibinfo {volume} {24}},\ \bibinfo {pages} {20904} (\bibinfo {year} {2016})}\BibitemShut {NoStop}%
\bibitem [{\citenamefont {Reitzenstein}\ and\ \citenamefont {Forchel}(2010)}]{reitzenstein_quantum_2010}%
  \BibitemOpen
  \bibfield  {author} {\bibinfo {author} {\bibfnamefont {S.}~\bibnamefont {Reitzenstein}}\ and\ \bibinfo {author} {\bibfnamefont {A.}~\bibnamefont {Forchel}},\ }\bibfield  {title} {\bibinfo {title} {Quantum dot micropillars},\ }\href {https://doi.org/10.1088/0022-3727/43/3/033001} {\bibfield  {journal} {\bibinfo  {journal} {J. Phys. D: Appl. Phys.}\ }\textbf {\bibinfo {volume} {43}},\ \bibinfo {pages} {033001} (\bibinfo {year} {2010})}\BibitemShut {NoStop}%
\bibitem [{\citenamefont {Ding}\ \emph {et~al.}(2016)\citenamefont {Ding}, \citenamefont {He}, \citenamefont {Duan}, \citenamefont {Gregersen}, \citenamefont {Chen}, \citenamefont {Unsleber}, \citenamefont {Maier}, \citenamefont {Schneider}, \citenamefont {Kamp}, \citenamefont {Höfling}, \citenamefont {Lu},\ and\ \citenamefont {Pan}}]{ding_-demand_2016}%
  \BibitemOpen
  \bibfield  {author} {\bibinfo {author} {\bibfnamefont {X.}~\bibnamefont {Ding}}, \bibinfo {author} {\bibfnamefont {Y.}~\bibnamefont {He}}, \bibinfo {author} {\bibfnamefont {Z.-C.}\ \bibnamefont {Duan}}, \bibinfo {author} {\bibfnamefont {N.}~\bibnamefont {Gregersen}}, \bibinfo {author} {\bibfnamefont {M.-C.}\ \bibnamefont {Chen}}, \bibinfo {author} {\bibfnamefont {S.}~\bibnamefont {Unsleber}}, \bibinfo {author} {\bibfnamefont {S.}~\bibnamefont {Maier}}, \bibinfo {author} {\bibfnamefont {C.}~\bibnamefont {Schneider}}, \bibinfo {author} {\bibfnamefont {M.}~\bibnamefont {Kamp}}, \bibinfo {author} {\bibfnamefont {S.}~\bibnamefont {Höfling}}, \bibinfo {author} {\bibfnamefont {C.-Y.}\ \bibnamefont {Lu}},\ and\ \bibinfo {author} {\bibfnamefont {J.-W.}\ \bibnamefont {Pan}},\ }\bibfield  {title} {{\selectlanguage {en}\bibinfo {title} {On-{Demand} {Single} {Photons} with {High} {Extraction} {Efficiency} and {Near}-{Unity} {Indistinguishability} from a {Resonantly} {Driven} {Quantum} {Dot} in a {Micropillar}}},\
  }\href {https://doi.org/10.1103/PhysRevLett.116.020401} {\bibfield  {journal} {\bibinfo  {journal} {Phys. Rev. Lett.}\ }\textbf {\bibinfo {volume} {116}},\ \bibinfo {pages} {020401} (\bibinfo {year} {2016})}\BibitemShut {NoStop}%
\bibitem [{\citenamefont {Burns}\ \emph {et~al.}(1977)\citenamefont {Burns}, \citenamefont {Milton},\ and\ \citenamefont {Lee}}]{burns_optical_1977}%
  \BibitemOpen
  \bibfield  {author} {\bibinfo {author} {\bibfnamefont {W.~K.}\ \bibnamefont {Burns}}, \bibinfo {author} {\bibfnamefont {A.~F.}\ \bibnamefont {Milton}},\ and\ \bibinfo {author} {\bibfnamefont {A.~B.}\ \bibnamefont {Lee}},\ }\bibfield  {title} {{\selectlanguage {en}\bibinfo {title} {Optical waveguide parabolic coupling horns}},\ }\href {https://doi.org/10.1063/1.89199} {\bibfield  {journal} {\bibinfo  {journal} {Applied Physics Letters}\ }\textbf {\bibinfo {volume} {30}},\ \bibinfo {pages} {28} (\bibinfo {year} {1977})}\BibitemShut {NoStop}%
\bibitem [{\citenamefont {Artioli}\ \emph {et~al.}(2019)\citenamefont {Artioli}, \citenamefont {Kotal}, \citenamefont {Gregersen}, \citenamefont {Verlot}, \citenamefont {Gérard},\ and\ \citenamefont {Claudon}}]{artioli_design_2019}%
  \BibitemOpen
  \bibfield  {author} {\bibinfo {author} {\bibfnamefont {A.}~\bibnamefont {Artioli}}, \bibinfo {author} {\bibfnamefont {S.}~\bibnamefont {Kotal}}, \bibinfo {author} {\bibfnamefont {N.}~\bibnamefont {Gregersen}}, \bibinfo {author} {\bibfnamefont {P.}~\bibnamefont {Verlot}}, \bibinfo {author} {\bibfnamefont {J.-M.}\ \bibnamefont {Gérard}},\ and\ \bibinfo {author} {\bibfnamefont {J.}~\bibnamefont {Claudon}},\ }\bibfield  {title} {{\selectlanguage {en}\bibinfo {title} {Design of {Quantum} {Dot}-{Nanowire} {Single}-{Photon} {Sources} that are {Immune} to {Thermomechanical} {Decoherence}}},\ }\href {https://doi.org/10.1103/PhysRevLett.123.247403} {\bibfield  {journal} {\bibinfo  {journal} {Phys. Rev. Lett.}\ }\textbf {\bibinfo {volume} {123}},\ \bibinfo {pages} {247403} (\bibinfo {year} {2019})}\BibitemShut {NoStop}%
\bibitem [{\citenamefont {Pearson}\ and\ \citenamefont {Faux}(2000)}]{pearson_analytical_2000}%
  \BibitemOpen
  \bibfield  {author} {\bibinfo {author} {\bibfnamefont {G.~S.}\ \bibnamefont {Pearson}}\ and\ \bibinfo {author} {\bibfnamefont {D.~A.}\ \bibnamefont {Faux}},\ }\bibfield  {title} {{\selectlanguage {en}\bibinfo {title} {Analytical solutions for strain in pyramidal quantum dots}},\ }\href {https://doi.org/10.1063/1.373729} {\bibfield  {journal} {\bibinfo  {journal} {Journal of Applied Physics}\ }\textbf {\bibinfo {volume} {88}},\ \bibinfo {pages} {730} (\bibinfo {year} {2000})}\BibitemShut {NoStop}%
\bibitem [{\citenamefont {Stepanov}\ \emph {et~al.}(2016)\citenamefont {Stepanov}, \citenamefont {Elzo-Aizarna}, \citenamefont {Bleuse}, \citenamefont {Malik}, \citenamefont {Curé}, \citenamefont {Gautier}, \citenamefont {Favre-Nicolin}, \citenamefont {Gérard},\ and\ \citenamefont {Claudon}}]{stepanov_large_2016}%
  \BibitemOpen
  \bibfield  {author} {\bibinfo {author} {\bibfnamefont {P.}~\bibnamefont {Stepanov}}, \bibinfo {author} {\bibfnamefont {M.}~\bibnamefont {Elzo-Aizarna}}, \bibinfo {author} {\bibfnamefont {J.}~\bibnamefont {Bleuse}}, \bibinfo {author} {\bibfnamefont {N.~S.}\ \bibnamefont {Malik}}, \bibinfo {author} {\bibfnamefont {Y.}~\bibnamefont {Curé}}, \bibinfo {author} {\bibfnamefont {E.}~\bibnamefont {Gautier}}, \bibinfo {author} {\bibfnamefont {V.}~\bibnamefont {Favre-Nicolin}}, \bibinfo {author} {\bibfnamefont {J.-M.}\ \bibnamefont {Gérard}},\ and\ \bibinfo {author} {\bibfnamefont {J.}~\bibnamefont {Claudon}},\ }\bibfield  {title} {{\selectlanguage {en}\bibinfo {title} {Large and {Uniform} {Optical} {Emission} {Shifts} in {Quantum} {Dots} {Strained} along {Their} {Growth} {Axis}}},\ }\href {https://doi.org/10.1021/acs.nanolett.6b00678} {\bibfield  {journal} {\bibinfo  {journal} {Nano Lett.}\ }\textbf {\bibinfo {volume} {16}},\ \bibinfo {pages} {3215} (\bibinfo {year} {2016})}\BibitemShut {NoStop}%
\bibitem [{\citenamefont {Vurgaftman}\ \emph {et~al.}(2001)\citenamefont {Vurgaftman}, \citenamefont {Meyer},\ and\ \citenamefont {Ram-Mohan}}]{vurgaftman_band_2001}%
  \BibitemOpen
  \bibfield  {author} {\bibinfo {author} {\bibfnamefont {I.}~\bibnamefont {Vurgaftman}}, \bibinfo {author} {\bibfnamefont {J.~R.}\ \bibnamefont {Meyer}},\ and\ \bibinfo {author} {\bibfnamefont {L.~R.}\ \bibnamefont {Ram-Mohan}},\ }\bibfield  {title} {{\selectlanguage {en}\bibinfo {title} {Band parameters for {III}–{V} compound semiconductors and their alloys}},\ }\href {https://doi.org/10.1063/1.1368156} {\bibfield  {journal} {\bibinfo  {journal} {Journal of Applied Physics}\ }\textbf {\bibinfo {volume} {89}},\ \bibinfo {pages} {5815} (\bibinfo {year} {2001})}\BibitemShut {NoStop}%
\bibitem [{\citenamefont {Ferreira~Neto}\ \emph {et~al.}(2024)\citenamefont {Ferreira~Neto}, \citenamefont {Bundgaard-Nielsen}, \citenamefont {Gregersen},\ and\ \citenamefont {Vannucci}}]{ferreira_neto_one-dimensional_2024}%
  \BibitemOpen
  \bibfield  {author} {\bibinfo {author} {\bibfnamefont {J.}~\bibnamefont {Ferreira~Neto}}, \bibinfo {author} {\bibfnamefont {M.}~\bibnamefont {Bundgaard-Nielsen}}, \bibinfo {author} {\bibfnamefont {N.}~\bibnamefont {Gregersen}},\ and\ \bibinfo {author} {\bibfnamefont {L.}~\bibnamefont {Vannucci}},\ }\bibfield  {title} {{\selectlanguage {en}\bibinfo {title} {One-dimensional photonic wire as a single-photon source: {Implications} of cavity {QED} to a phonon bath of reduced dimensionality}},\ }\href {https://doi.org/10.1103/PhysRevB.110.115308} {\bibfield  {journal} {\bibinfo  {journal} {Phys. Rev. B}\ }\textbf {\bibinfo {volume} {110}},\ \bibinfo {pages} {115308} (\bibinfo {year} {2024})}\BibitemShut {NoStop}%
\bibitem [{\citenamefont {Braakman}\ and\ \citenamefont {Poggio}(2019)}]{braakman_force_2019}%
  \BibitemOpen
  \bibfield  {author} {\bibinfo {author} {\bibfnamefont {F.~R.}\ \bibnamefont {Braakman}}\ and\ \bibinfo {author} {\bibfnamefont {M.}~\bibnamefont {Poggio}},\ }\bibfield  {title} {\bibinfo {title} {Force sensing with nanowire cantilevers},\ }\href {https://doi.org/10.1088/1361-6528/ab19cf} {\bibfield  {journal} {\bibinfo  {journal} {Nanotechnology}\ }\textbf {\bibinfo {volume} {30}},\ \bibinfo {pages} {332001} (\bibinfo {year} {2019})}\BibitemShut {NoStop}%
\bibitem [{\citenamefont {Lavrinenko}\ \emph {et~al.}(2018)\citenamefont {Lavrinenko}, \citenamefont {Lægsgaard}, \citenamefont {Gregersen}, \citenamefont {Schmidt},\ and\ \citenamefont {Søndergaard}}]{lavrinenko_numerical_2018}%
  \BibitemOpen
  \bibfield  {author} {\bibinfo {author} {\bibfnamefont {A.~V.}\ \bibnamefont {Lavrinenko}}, \bibinfo {author} {\bibfnamefont {J.}~\bibnamefont {Lægsgaard}}, \bibinfo {author} {\bibfnamefont {N.}~\bibnamefont {Gregersen}}, \bibinfo {author} {\bibfnamefont {F.}~\bibnamefont {Schmidt}},\ and\ \bibinfo {author} {\bibfnamefont {T.}~\bibnamefont {Søndergaard}},\ }\href {https://doi.org/10.1201/b17408} {{\selectlanguage {en}\emph {\bibinfo {title} {Numerical {Methods} in {Photonics}}}}},\ \bibinfo {edition} {1st}\ ed.\ (\bibinfo  {publisher} {CRC Press},\ \bibinfo {year} {2018})\BibitemShut {NoStop}%
\bibitem [{\citenamefont {Gür}\ \emph {et~al.}(2021)\citenamefont {Gür}, \citenamefont {Arslanagić}, \citenamefont {Mattes},\ and\ \citenamefont {Gregersen}}]{gur_open-geometry_2021}%
  \BibitemOpen
  \bibfield  {author} {\bibinfo {author} {\bibfnamefont {U.~M.}\ \bibnamefont {Gür}}, \bibinfo {author} {\bibfnamefont {S.}~\bibnamefont {Arslanagić}}, \bibinfo {author} {\bibfnamefont {M.}~\bibnamefont {Mattes}},\ and\ \bibinfo {author} {\bibfnamefont {N.}~\bibnamefont {Gregersen}},\ }\bibfield  {title} {{\selectlanguage {en}\bibinfo {title} {Open-geometry modal method based on transverse electric and transverse magnetic mode expansion for orthogonal curvilinear coordinates}},\ }\href {https://doi.org/10.1103/PhysRevE.103.033301} {\bibfield  {journal} {\bibinfo  {journal} {Phys. Rev. E}\ }\textbf {\bibinfo {volume} {103}},\ \bibinfo {pages} {033301} (\bibinfo {year} {2021})}\BibitemShut {NoStop}%
\bibitem [{\citenamefont {Munsch}\ \emph {et~al.}(2013{\natexlab{a}})\citenamefont {Munsch}, \citenamefont {Malik}, \citenamefont {Dupuy}, \citenamefont {Delga}, \citenamefont {Bleuse}, \citenamefont {Gérard}, \citenamefont {Claudon}, \citenamefont {Gregersen},\ and\ \citenamefont {Mørk}}]{munsch_dielectric_2013}%
  \BibitemOpen
  \bibfield  {author} {\bibinfo {author} {\bibfnamefont {M.}~\bibnamefont {Munsch}}, \bibinfo {author} {\bibfnamefont {N.~S.}\ \bibnamefont {Malik}}, \bibinfo {author} {\bibfnamefont {E.}~\bibnamefont {Dupuy}}, \bibinfo {author} {\bibfnamefont {A.}~\bibnamefont {Delga}}, \bibinfo {author} {\bibfnamefont {J.}~\bibnamefont {Bleuse}}, \bibinfo {author} {\bibfnamefont {J.-M.}\ \bibnamefont {Gérard}}, \bibinfo {author} {\bibfnamefont {J.}~\bibnamefont {Claudon}}, \bibinfo {author} {\bibfnamefont {N.}~\bibnamefont {Gregersen}},\ and\ \bibinfo {author} {\bibfnamefont {J.}~\bibnamefont {Mørk}},\ }\bibfield  {title} {{\selectlanguage {en}\bibinfo {title} {Dielectric {GaAs} {Antenna} {Ensuring} an {Efficient} {Broadband} {Coupling} between an {InAs} {Quantum} {Dot} and a {Gaussian} {Optical} {Beam}}},\ }\href {https://doi.org/10.1103/PhysRevLett.110.177402} {\bibfield  {journal} {\bibinfo  {journal} {Phys. Rev. Lett.}\ }\textbf {\bibinfo {volume} {110}},\ \bibinfo {pages} {177402} (\bibinfo {year}
  {2013}{\natexlab{a}})}\BibitemShut {NoStop}%
\bibitem [{\citenamefont {Munsch}\ \emph {et~al.}(2013{\natexlab{b}})\citenamefont {Munsch}, \citenamefont {Malik}, \citenamefont {Dupuy}, \citenamefont {Delga}, \citenamefont {Bleuse}, \citenamefont {Gérard}, \citenamefont {Claudon}, \citenamefont {Gregersen},\ and\ \citenamefont {Mørk}}]{munsch_erratum_2013}%
  \BibitemOpen
  \bibfield  {author} {\bibinfo {author} {\bibfnamefont {M.}~\bibnamefont {Munsch}}, \bibinfo {author} {\bibfnamefont {N.~S.}\ \bibnamefont {Malik}}, \bibinfo {author} {\bibfnamefont {E.}~\bibnamefont {Dupuy}}, \bibinfo {author} {\bibfnamefont {A.}~\bibnamefont {Delga}}, \bibinfo {author} {\bibfnamefont {J.}~\bibnamefont {Bleuse}}, \bibinfo {author} {\bibfnamefont {J.-M.}\ \bibnamefont {Gérard}}, \bibinfo {author} {\bibfnamefont {J.}~\bibnamefont {Claudon}}, \bibinfo {author} {\bibfnamefont {N.}~\bibnamefont {Gregersen}},\ and\ \bibinfo {author} {\bibfnamefont {J.}~\bibnamefont {Mørk}},\ }\bibfield  {title} {{\selectlanguage {en}\bibinfo {title} {Erratum: {Dielectric} {GaAs} {Antenna} {Ensuring} an {Efficient} {Broadband} {Coupling} between an {InAs} {Quantum} {Dot} and a {Gaussian} {Optical} {Beam} [{Phys}. {Rev}. {Lett}. \textbf{110} , 177402 (2013)]}},\ }\href {https://doi.org/10.1103/PhysRevLett.111.239902} {\bibfield  {journal} {\bibinfo  {journal} {Phys. Rev. Lett.}\ }\textbf {\bibinfo {volume}
  {111}},\ \bibinfo {pages} {239902} (\bibinfo {year} {2013}{\natexlab{b}})}\BibitemShut {NoStop}%
\bibitem [{\citenamefont {Nazir}\ and\ \citenamefont {McCutcheon}(2016)}]{nazir_modelling_2016}%
  \BibitemOpen
  \bibfield  {author} {\bibinfo {author} {\bibfnamefont {A.}~\bibnamefont {Nazir}}\ and\ \bibinfo {author} {\bibfnamefont {D.~P.~S.}\ \bibnamefont {McCutcheon}},\ }\bibfield  {title} {\bibinfo {title} {Modelling exciton–phonon interactions in optically driven quantum dots},\ }\href {https://doi.org/10.1088/0953-8984/28/10/103002} {\bibfield  {journal} {\bibinfo  {journal} {J. Phys.: Condens. Matter}\ }\textbf {\bibinfo {volume} {28}},\ \bibinfo {pages} {103002} (\bibinfo {year} {2016})}\BibitemShut {NoStop}%
\end{thebibliography}%
\newpage
\onecolumngrid
\section*{Supplementary Material}

\onecolumngrid

\renewcommand{\thetable}{\Alph{table}}
\setcounter{table}{0} 

\renewcommand{\thefigure}{\Alph{figure}}
\setcounter{figure}{0} 

\renewcommand{\theequation}{S\arabic{equation}}
\setcounter{equation}{0} 

\begin{center}
\textbf{Thermally induced vibration modes}
\end{center}

\begin{figure}[H]
\centering
\includegraphics[width=0.60\linewidth]{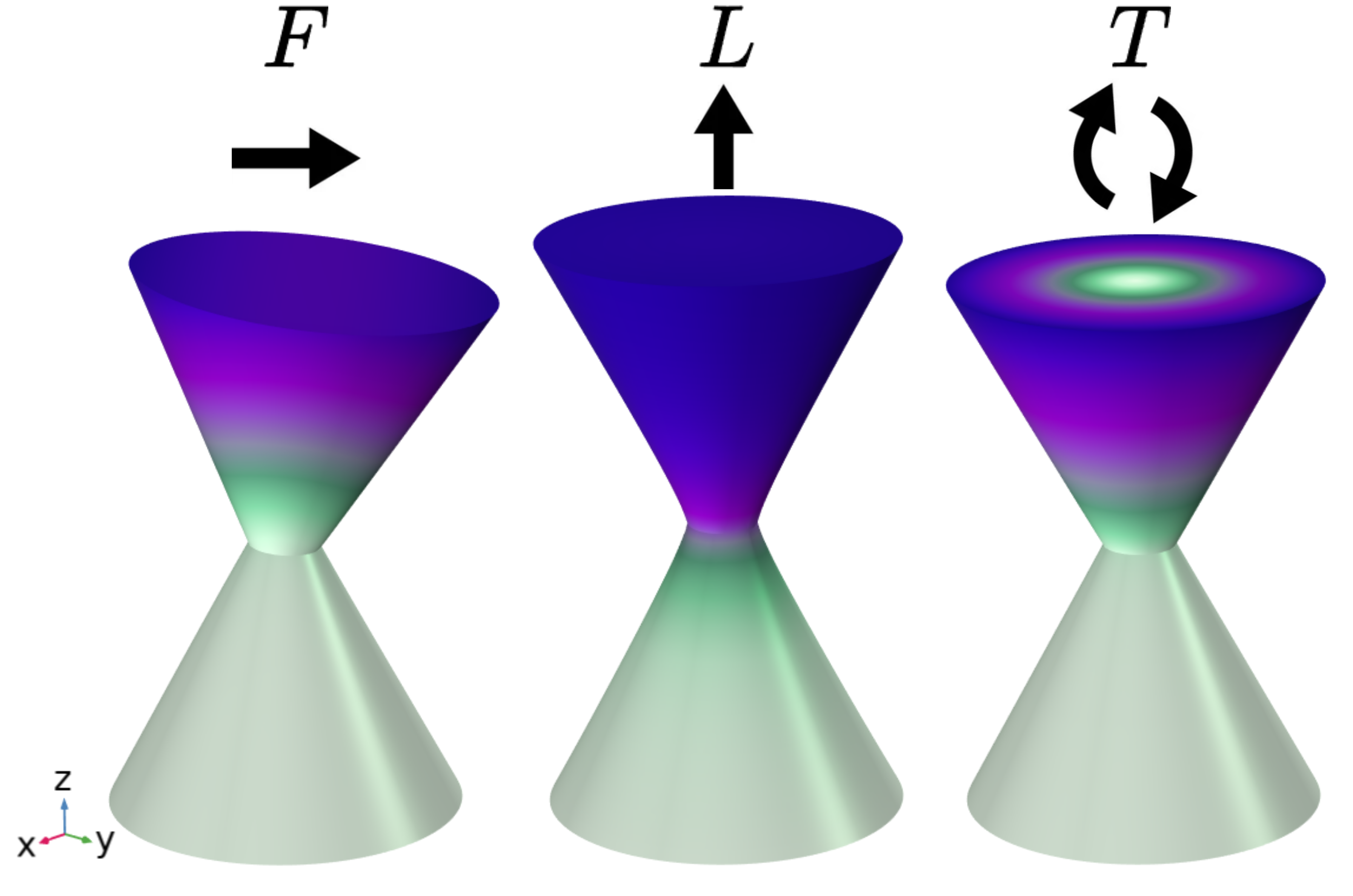}
\caption{Representation of various vibration modes for each family, shown in terms of the mode displacement. The color bar is omitted for compactness, with white/light green indicating lower values and dark blue representing peak displacement. Above, the dimensions of the photonic hourglass structure are not up to scale and have been exaggerated for improved clarity.}
\label{fig:hourglass_modes}
\end{figure}

Given the complexity of the displacement profile for the mechanical modes, we simulate the mechanical resonator using a finite element method (FEM) solver. We conduct an eigenmode study where we apply a fixed boundary condition to one of the ends of our geometry. As mentioned in the main text of the manuscript, the vibrational properties, including the mechanical eigenfrequencies $\omega_{m}$ for the free-standing geometry (singly clamped to a substrate), their effective mass $m^{\mathrm{eff}}_{m}$, and primary strain tensors ($\epsilon_{xx}$, $\epsilon_{yy}$, $\epsilon_{zz}$), are calculated with the FEM software. The strain tensors are extracted at points along the beam waist cross-section defined by the direction of maximum displacement at a given height from the substrate. The retrieved modes are then classified as flexural ($F$), longitudinal ($L$), or torsional ($T$), as shown in Fig.~\ref{fig:hourglass_modes}, so that we can properly account for their contributions. 

The mechanical resonator is described by a harmonic oscillator via

\begin{equation}
\label{eqn:fem_1}
u(r,t)=x(t)|u(r)|,
\end{equation}

where $|u(r)| = u(r)/\mathrm{max}(|u(r)|)$ is the normalized displacement obtained from the FEM solver. The time-dependent displacement $x(t)$ is defined by the equation of motion of an effective 1D harmonic oscillator 

\begin{equation}
\label{eqn:fem_2}
m^{\mathrm{eff}}_{m}\frac{\mathrm{d}^2 x(t)}{\mathrm{d} t^2}+m^{\mathrm{eff}}_{m}\Gamma_{m}\frac{\mathrm{d} x(t)}{\mathrm{d} t}+m^{\mathrm{eff}}_{m}\Omega_{m}^2x(t)=F(t),
\end{equation}

where $\Omega_{m}/2\pi$ is the mechanical frequency, $k^{\mathrm{eff}}_{m}=m^{\mathrm{eff}}_{m}\Omega_{m}^{2}$ the
spring constant, and $\Gamma_{m}$ the energy dissipation rate, related
to the mechanical quality factor by $Q_{m}=\Omega_{m}/\Gamma_{m}$. Here, the effective, or motional, mass is obtained via thermomechanical calibration by performing a volume integration of the normalized displacement
\begin{equation}
\label{eqn:fem_3}
m^{\mathrm{eff}}_{m}=\int\rho_{m}(r)|u(r)|^{2}\mathrm{d}V,
\end{equation}
where $\rho_{m}$ is the material density. 

We recall that the modulation of the QD bandgap via deformation potential is given by 
\begin{equation}
\label{eqn:fem_7}
\hbar\frac{\partial \omega_{\mathrm{eg}}}{\partial u_{m}}=a\left(\frac{\partial \epsilon_{\mathrm{h}}}{\partial u_{m}}\right)+\frac{b}{2}\left(\frac{\partial \epsilon_{\mathrm{sh}}}{\partial u_{m}}\right),
\end{equation}
which corresponds to Eq. (\textcolor{blue}{2}) of the main text, with all relevant quantities formerly defined therein.

Following, we evaluate the optomechanical coupling strength 
\begin{equation}
\label{eqn:fem_5}
g_{m}=\left(\frac{\partial \omega_{\mathrm{eg}}}{\partial u_{m}}\right)u^{\mathrm{zpf}}_{m}=\left(a\partial_u\epsilon_{\mathrm{h}}+\frac{b}{2}\partial_u\epsilon_{\mathrm{sh}}\right)u^{\mathrm{zpf}}_{m},
\end{equation}
where $u^{\textrm{zpf}}_{m}=\sqrt{\hbar/(2m^{\textrm{eff}}_{m}\Omega_{m})}$ stands for the root mean square (RMS) value of the zero-point fluctuation of the mode displacement. 

\begin{table}[hbtp]
\centering
\begin{tabular}{cccc}
\hline \hline
\addlinespace[0.25em]
 & GaAs & AlGaAs & AR coating \\ \addlinespace[0.125em] \hline
\addlinespace[0.125em]
Mass density, $\rho_{m}$ [kg/m$^3$] & 5316 & 5316 & 5660 \\ 
Young modulus, $Y$ [GPa] & 85.9 & 77.71 & 140 \\ 
Poisson ratio, $\nu$ & 0.31 & 0.31 & 0.358 \\ \hline \hline
\end{tabular}
\caption{Material properties used in the study of mechanical vibrations with FEM simulations.}
\label{tab:material_properties}
\end{table}

Table~\ref{tab:material_properties} presents the material properties used in our FEM simulations, whereas Table~\ref{tab:SI_without} (\ref{tab:SI_with}) displays the extracted results for a photonic hourglass without (with) DBRs. We emphasize that $g_{m}^{\mathrm{max}}$ is obtained at the sidewall position of the structure beam waist diameter for flexural modes and at the center for longitudinal modes, corresponding to the locations where each mode reaches its maximum strain profile $\epsilon_{zz}$.

Lastly, Tables~\ref{tab:SI_couplings_without} and \ref{tab:SI_couplings_with} present the calculated coupling parameters $\theta^{2}_{m}$ and $\eta^{2}_{m}$ for the geometries without and with DBRs, respectively.

\newpage

\begin{table}[H]
\centering
\begin{minipage}{0.45\textwidth}
\centering
\begin{tabular}{ccccccc}
\hline \hline 
\addlinespace[0.25em]
Mode                             & $\Omega_{m}/2\pi$ {[}MHz{]} & $m^{\mathrm{eff}}_{m}$ {[}pg{]} & $u^{\mathrm{zpf}}_{m}$ {[}fm{]} & $|g_{m}^{\mathrm{max}}|/2\pi$ {[}kHz{]} \\
\addlinespace[0.125em]
\midrule
$F_{1}$ & 0.0091                  & 182.6                  & 70.9                  & 34.19 \\
$F_{2}$ & 0.2306                & 327.8                   & 10.5                   & 32.63 \\
$F_{3}$ & 0.7581                  & 2130.5                   & 2.3                   & 42.42 \\
$F_{4}$ & 1.51                  & 594.6                   & 3.1                   & 132.70 \\
$F_{5}$ & 2.19                & 813.8                   & 2.2                   & 57.06 \\
$F_{6}$ & 3.30                  & 632.3                   & 2.0                   & 237.57 \\
$F_{7}$ & 4.38                  & 691.9                   & 1.7                   & 60.42 \\
$F_{8}$ & 5.84                & 634.7                   & 1.5                   & 336.98 \\
$F_{9}$ & 7.32                  & 654.2                   & 1.3                   & 62.55 \\
$F_{10}$ & 9.13                  & 629.7                   & 1.2                   & 434.58 \\
$F_{11}$ & 11.00                & 635.8                   & 1.1                   & 63.69 \\
$F_{12}$ & 13.17                  & 622.9                   & 1.0                   & 529.99 \\
$F_{13}$ & 15.43                  & 623.5                   & 0.9                   & 65.09 \\
$F_{14}$ & 17.94                & 615.7                   & 0.9                   & 621.44 \\
$F_{15}$ & 20.58                  & 613.7                   & 0.8                   & 65.84 \\
$F_{16}$ & 23.44                  & 608.3                   & 0.8                   & 719.45 \\
$F_{17}$ & 26.44                & 604.8                   & 0.7                   & 66.01 \\
$F_{18}$ & 29.65                  & 600.6                   & 0.7                   & 804.47 \\
$F_{19}$ & 33.01                  & 596.4                   & 0.7                   & 65.50 \\
$F_{20}$ & 36.55                & 592.6                   & 0.6                   & 889.28 \\
$F_{21}$ & 40.25                  & 588.0                   & 0.6                   & 64.42 \\
$F_{22}$ & 44.12                  & 584.3                   & 0.6                   & 979.83 \\
$F_{23}$ & 48.15                & 579.5                   & 0.5                   & 62.19 \\
$F_{24}$ & 52.34                  & 576.0                   & 0.5                   & 1060.20 \\
$F_{25}$ & 56.70                & 571.1                   & 0.5                   & 59.26 \\
$F_{26}$ & 61.19                  & 567.4                   & 0.5                   & 1145.10 \\
$F_{27}$ & 65.86                  & 562.5                   & 0.4                   & 55.02 \\
$F_{28}$ & 70.65                & 558.8                   & 0.4                   & 1237.23 \\
$F_{29}$ & 75.62                  & 553.9                   & 0.4                   & 50.38 \\
$F_{30}$ & 80.70                  & 549.9                   & 0.4                   & 1311.12 \\
$F_{31}$ & 85.96                & 545.0                   & 0.4                   & 44.22 \\
$F_{32}$ & 91.32                  & 541.2                   & 0.4                   & 1394.07 \\
$F_{33}$ & 96.85                  & 536.1                   & 0.4                   & 36.56 \\
$F_{34}$ & 102.48                & 532.2                   & 0.3                   & 1463.32 \\
$F_{35}$ & 108.28                  & 527.4                   & 0.3                   & 28.82 \\
$F_{36}$ & 114.16                  & 523.2                   & 0.3                   & 1531.23 \\
$F_{37}$ & 120.21                & 518.8                   & 0.3                   & 18.70 \\
$F_{38}$ & 126.34                  & 514.3                   & 0.3                   & 1610.11 \\
$F_{39}$ & 132.64                  & 509.8                   & 0.3                   & 8.81 \\
$F_{40}$ & 139.00                & 505.3                   & 0.3                   & 1682.01 \\
\toprule 
$L_{1}$ & 0.6954                       & 97.1                   & 11.2                  & 1.73 \\
$L_{2}$ & 4.49                      & 311.0                   & 2.5                   & 241.79 \\
$L_{3}$ & 26.79                    & 383275.4                   & 0.0286                   & 0.62 \\
$L_{4}$ & 32.13                    & 11978.2                   & 0.1                   & 202.18 \\
$L_{5}$ & 34.42                    & 237.6                   & 1.0                   & 1.54 \\
$L_{6}$ & 46.42                    & 27755.6                   & 0.1                   & 1.14 \\
$L_{7}$ & 46.44                    & 478.7                   & 0.6                   & 282.10 \\
$L_{8}$ & 54.81                    & 236.4                   & 0.8                   & 1.45 \\
$L_{9}$ & 64.78                    & 3862.3                   & 0.2                   & 290.94 \\
$L_{10}$ & 66.08                    & 6963.3                   & 0.1                   & 1.60 \\
$L_{11}$ & 74.96                    & 240.0                   & 0.7                   & 1.92 \\
$L_{12}$ & 80.27                    & 516.2                   & 0.5                   & 363.59 \\
$L_{13}$ & 86.00                    & 3103.1                   & 0.2                   & 3.66 \\
$L_{14}$ & 95.25                    & 247.6                   & 0.2                   & 4.00 \\
$L_{15}$ & 97.98                    & 2317.8                   & 0.4                   & 362.53 \\
$L_{16}$ & 106.19                    & 1904.6                   & 0.5                   & 2.22 \\
$L_{17}$ & 114.02                    & 552.7                   & 0.2                   & 432.42 \\
$L_{18}$ & 115.71                    & 257.4                   & 0.2                   & 4.76 \\
$L_{19}$ & 126.61                    & 1382.5                   & 0.5                   & 4.46 \\
$L_{20}$ & 131.55                    & 1759.9                   & 0.2                   & 423.81 \\
$L_{21}$ & 136.33                    & 267.7                   & 0.2 & 2.86 \\ \hline \hline
\end{tabular}
\caption{FEM of vibration modes for the photonic hourglass without DBRs.}
\label{tab:SI_without}
\end{minipage}\hfill
\begin{minipage}{0.45\textwidth}
\centering
\begin{tabular}{ccccccc}
\hline \hline 
\addlinespace[0.25em]
Mode                             & $\Omega_{m}/2\pi$ {[}MHz{]} & $m^{\mathrm{eff}}_{m}$ {[}pg{]} & $u^{\mathrm{zpf}}_{m}$ {[}fm{]} & $|g_{m}^{\mathrm{max}}|/2\pi$ {[}kHz{]} \\
\addlinespace[0.125em]
\midrule
$F_{1}$ & 0.0091                  & 182.6                  & 70.9                  & 34.19 \\
$F_{2}$ & 0.2301                & 328.0                   & 10.5                   & 32.77 \\
$F_{3}$ & 0.7572                  & 2112.7                   & 2.3                   & 42.60 \\
$F_{4}$ & 1.51                  & 599.5                   & 3.0                   & 132.15 \\
$F_{5}$ & 2.18                & 815.9                   & 2.2                   & 54.92 \\
$F_{6}$ & 3.30                  & 630.6                   & 2.0                   & 236.73 \\
$F_{7}$ & 4.37                  & 686.9                   & 1.7                   & 60.34 \\
$F_{8}$ & 5.83                & 638.5                   & 1.5                   & 335.50 \\
$F_{9}$ & 7.29                  & 660.7                   & 1.3                   & 57.26 \\
$F_{10}$ & 9.11                  & 630.9                   & 1.2                   & 433.78 \\
$F_{11}$ & 10.99                & 628.3                   & 1.1                   & 60.34 \\
$F_{12}$ & 13.14                  & 620.9                   & 1.0                   & 528.17 \\
$F_{13}$ & 15.39                  & 631.1                   & 0.9                   & 59.31 \\
$F_{14}$ & 17.90                & 622.6                   & 0.9                   & 622.36 \\
$F_{15}$ & 20.54                  & 609.1                   & 0.8                   & 53.52 \\
$F_{16}$ & 23.40                  & 601.5                   & 0.8                   & 719.03 \\
$F_{17}$ & 26.39                & 607.7                   & 0.7                   & 58.08 \\
$F_{18}$ & 29.59                  & 609.6                   & 0.7                   & 808.20 \\
$F_{19}$ & 32.96                  & 597.4                   & 0.7                   & 47.50 \\
$F_{20}$ & 36.48                & 585.8                   & 0.6                   & 886.38 \\
$F_{21}$ & 40.15                  & 586.6                   & 0.6                   & 47.68 \\
$F_{22}$ & 44.02                  & 591.7                   & 0.6                   & 976.90 \\
$F_{23}$ & 48.08                & 583.5                   & 0.5                   & 39.79 \\
$F_{24}$ & 52.26                  & 570.9                   & 0.5                   & 1059.08 \\
$F_{25}$ & 56.57                & 568.1                   & 0.5                   & 34.17 \\
$F_{26}$ & 61.04                  & 573.5                   & 0.5                   & 1146.89 \\
$F_{27}$ & 65.75                  & 568.3                   & 0.4                   & 21.76 \\
$F_{28}$ & 70.55                & 554.3                   & 0.4                   & 1234.79 \\
$F_{29}$ & 75.48                  & 548.6                   & 0.4                   & 20.11 \\
$F_{30}$ & 80.51                  & 554.8                   & 0.4                   & 1309.93 \\
$F_{31}$ & 85.79                & 553.5                   & 0.4                   & 3.23 \\
$F_{32}$ & 91.17                  & 538.9                   & 0.4                   & 1392.81 \\
$F_{33}$ & 96.67                  & 528.7                   & 0.4                   & 5.07 \\
$F_{34}$ & 102.26                & 533.6                   & 0.3                   & 1461.22 \\
$F_{35}$ & 108.06                  & 537.4                   & 0.3                   & 24.80 \\
$F_{36}$ & 113.96                  & 524.4                   & 0.3                   & 1534.89 \\
$F_{37}$ & 119.98                & 511.8                   & 0.3                   & 43.02 \\
$F_{38}$ & 126.07                  & 511.8                   & 0.3                   & 1606.96 \\
$F_{39}$ & 132.38                  & 517.9                   & 0.3                   & 50.62 \\
$F_{40}$ & 138.76                & 510.0                   & 0.3                   & 1674.53 \\
\toprule
$L_{1}$ & 0.6953                       & 97.1                   & 11.2                  & 1.73 \\
$L_{2}$ & 4.49                      & 311.0                   & 2.5                   & 241.34 \\
$L_{3}$ & 26.73                    & 383275.4                   & 0.0286                   & 0.63 \\
$L_{4}$ & 32.06                    & 11978.2                   & 0.1                   & 203.20 \\
$L_{5}$ & 34.38                    & 238.7                   & 1.0                   & 1.54 \\
$L_{6}$ & 46.09                    & 29345.7                   & 0.1                   & 1.13 \\
$L_{7}$ & 46.39                    & 481.0                   & 0.6                   & 282.35 \\
$L_{8}$ & 54.72                    & 235.5                   & 0.8                   & 1.44 \\
$L_{9}$ & 64.42                    & 3968.9                   & 0.2                   & 291.95 \\
$L_{10}$ & 66.04                    & 7115.4                   & 0.1                   & 1.58 \\
$L_{11}$ & 74.94                    & 240.9                   & 0.7                   & 1.91 \\
$L_{12}$ & 80.14                    & 512.2                   & 0.5                   & 359.37 \\
$L_{13}$ & 85.55                    & 3129.1                   & 0.2                   & 1.99 \\
$L_{14}$ & 95.09                    & 247.6                   & 0.2                   & 3.98 \\
$L_{15}$ & 97.89                    & 2324.6                   & 0.4                   & 363.84 \\
$L_{16}$ & 106.02                    & 1984.3                   & 0.5                   & 2.22 \\
$L_{17}$ & 113.94                    & 559.7                   & 0.2                   & 429.80 \\
$L_{18}$ & 115.63                    & 256.3                   & 0.2                   & 4.62 \\
$L_{19}$ & 126.12                    & 1362.4                   & 0.5                   & 4.90 \\
$L_{20}$ & 130.98                    & 1756.5                   & 0.2                   & 432.31 \\
$L_{21}$ & 136.13                    & 270.8                   & 0.2                   & 2.83 \\ \hline \hline
\end{tabular}
\caption{FEM of vibration modes for the photonic hourglass with DBRs.}
\label{tab:SI_with}
\end{minipage}
\end{table}

\newpage

\begin{table}[H]
\centering
\begin{minipage}{0.45\textwidth}
\centering
\begin{tabular}{cccccc}
\hline \hline 
\addlinespace[0.25em]
Mode                             & $\Omega_{m}/2\pi$ {[}MHz{]} & $\theta^{2}_{m}$ & $\eta^{2}_{m}$ \\
\addlinespace[0.125em]
\midrule
$F_{1}$  & 0.0091  & 5.11$\times10^{8}$  & 1.40$\times10^{1}$ \\
$F_{2}$  & 0.2306  & 2.89$\times10^{4}$  & 2.00$\times10^{-2}$ \\
$F_{3}$  & 0.7581  & 1.38$\times10^{3}$  & 3.13$\times10^{-3}$ \\
$F_{4}$  & 1.51    & 1.70$\times10^{3}$  & 7.70$\times10^{-3}$ \\
$F_{5}$  & 2.19    & 1.04$\times10^{2}$  & 6.80$\times10^{-4}$ \\
$F_{6}$  & 3.30    & 5.22$\times10^{2}$  & 5.17$\times10^{-3}$ \\
$F_{7}$  & 4.38    & 1.45$\times10^{1}$  & 1.91$\times10^{-4}$ \\
$F_{8}$  & 5.84    & 1.90$\times10^{2}$  & 3.33$\times10^{-3}$ \\
$F_{9}$  & 7.32    & 3.33$\times10^{0}$  & 7.31$\times10^{-5}$ \\
$F_{10}$ & 9.13    & 8.28$\times10^{1}$  & 2.27$\times10^{-3}$ \\
$F_{11}$ & 11.00   & 1.02$\times10^{0}$  & 3.35$\times10^{-5}$ \\
$F_{12}$ & 13.17   & 4.10$\times10^{1}$  & 1.62$\times10^{-3}$ \\
$F_{13}$ & 15.43   & 3.85$\times10^{-1}$ & 1.78$\times10^{-5}$ \\
$F_{14}$ & 17.94   & 2.23$\times10^{1}$  & 1.20$\times10^{-3}$ \\
$F_{15}$ & 20.58   & 1.66$\times10^{-1}$ & 1.02$\times10^{-5}$ \\
$F_{16}$ & 23.44   & 1.34$\times10^{1}$  & 9.42$\times10^{-4}$ \\
$F_{17}$ & 26.44   & 7.86$\times10^{-2}$ & 6.23$\times10^{-6}$ \\
$F_{18}$ & 29.65   & 8.28$\times10^{0}$  & 7.36$\times10^{-4}$ \\
$F_{19}$ & 33.01   & 3.98$\times10^{-2}$ & 3.94$\times10^{-6}$ \\
$F_{20}$ & 36.55   & 5.40$\times10^{0}$  & 5.92$\times10^{-4}$ \\
$F_{21}$ & 40.25   & 2.12$\times10^{-2}$ & 2.56$\times10^{-6}$ \\
$F_{22}$ & 44.12   & 3.73$\times10^{0}$  & 4.93$\times10^{-4}$ \\
$F_{23}$ & 48.15   & 1.15$\times10^{-2}$ & 1.67$\times10^{-6}$ \\
$F_{24}$ & 52.34   & 2.61$\times10^{0}$  & 4.10$\times10^{-4}$ \\
$F_{25}$ & 56.70   & 6.42$\times10^{-3}$ & 1.09$\times10^{-6}$ \\
$F_{26}$ & 61.19   & 1.91$\times10^{0}$  & 3.50$\times10^{-4}$ \\
$F_{27}$ & 65.86   & 3.53$\times10^{-3}$ & 6.98$\times10^{-7}$ \\
$F_{28}$ & 70.65   & 1.45$\times10^{0}$  & 3.07$\times10^{-4}$ \\
$F_{29}$ & 75.62   & 1.96$\times10^{-3}$ & 4.44$\times10^{-7}$ \\
$F_{30}$ & 80.70   & 1.09$\times10^{0}$  & 2.64$\times10^{-4}$ \\
$F_{31}$ & 85.96   & 1.03$\times10^{-3}$ & 2.65$\times10^{-7}$ \\
$F_{32}$ & 91.32   & 8.50$\times10^{-1}$ & 2.33$\times10^{-4}$ \\
$F_{33}$ & 96.85   & 4.90$\times10^{-4}$ & 1.42$\times10^{-7}$ \\
$F_{34}$ & 102.48  & 6.63$\times10^{-1}$ & 2.04$\times10^{-4}$ \\
$F_{35}$ & 108.28  & 2.18$\times10^{-4}$ & 7.08$\times10^{-8}$ \\
$F_{36}$ & 114.16  & 5.25$\times10^{-1}$ & 1.62$\times10^{-4}$ \\
$F_{37}$ & 120.21  & 6.71$\times10^{-5}$ & 4.42$\times10^{-9}$ \\
$F_{38}$ & 126.34  & 4.28$\times10^{-1}$ & 1.46$\times10^{-4}$ \\
$F_{39}$ & 132.64  & 1.11$\times10^{-5}$ & 2.00$\times10^{-5}$ \\
$F_{40}$ & 139.00  & 3.51$\times10^{-1}$ & 1.46$\times10^{-4}$ \\
\toprule 
$L_{1}$  & 0.6954   & 2.97$\times10^{0}$     & 6.19$\times10^{-6}$ \\
$L_{2}$  & 4.49     & 2.15$\times10^{2}$     & 2.90$\times10^{-3}$ \\
$L_{3}$  & 26.79    & 6.75$\times10^{-6}$    & 5.42$\times10^{-10}$ \\
$L_{4}$  & 32.13    & 4.11$\times10^{-1}$    & 3.96$\times10^{-5}$ \\
$L_{5}$  & 34.42    & 1.93$\times10^{-5}$    & 1.99$\times10^{-9}$ \\
$L_{6}$  & 46.42    & 4.33$\times10^{-6}$    & 6.02$\times10^{-10}$ \\
$L_{7}$  & 46.44    & 2.65$\times10^{-1}$    & 3.69$\times10^{-5}$ \\
$L_{8}$  & 54.81    & 4.26$\times10^{-6}$    & 7.00$\times10^{-10}$ \\
$L_{9}$  & 64.78    & 1.04$\times10^{-1}$    & 2.02$\times10^{-5}$ \\
$L_{10}$ & 66.08    & 2.95$\times10^{-6}$    & 5.84$\times10^{-10}$ \\
$L_{11}$ & 74.96    & 2.93$\times10^{-6}$    & 6.59$\times10^{-10}$ \\
$L_{12}$ & 80.27    & 8.52$\times10^{-2}$    & 2.05$\times10^{-5}$ \\
$L_{13}$ & 86.00    & 7.00$\times10^{-6}$    & 1.81$\times10^{-9}$ \\
$L_{14}$ & 95.25    & 6.18$\times10^{-6}$    & 1.77$\times10^{-9}$ \\
$L_{15}$ & 97.98    & 4.66$\times10^{-2}$    & 1.37$\times10^{-5}$ \\
$L_{16}$ & 106.19   & 1.37$\times10^{-6}$    & 4.38$\times10^{-10}$ \\
$L_{17}$ & 114.02   & 4.20$\times10^{-2}$    & 1.44$\times10^{-5}$ \\
$L_{18}$ & 115.71   & 4.87$\times10^{-6}$    & 1.69$\times10^{-9}$ \\
$L_{19}$ & 126.61   & 3.26$\times10^{-6}$    & 1.24$\times10^{-9}$ \\
$L_{20}$ & 131.55   & 2.63$\times10^{-2}$    & 1.04$\times10^{-5}$ \\
$L_{21}$ & 136.33   & 1.07$\times10^{-6}$    & 4.40$\times10^{-10}$ \\
\hline \hline
\end{tabular}
\caption{Extracted parameters $\theta^{2}_{m}$ and $\eta^{2}_{m}$ for the geometry without DBRs.}
\label{tab:SI_couplings_without}
\end{minipage}\hfill
\begin{minipage}{0.45\textwidth}
\centering
\begin{tabular}{cccccc}
\hline \hline 
\addlinespace[0.25em]
Mode                             & $\Omega_{m}/2\pi$ {[}MHz{]} & $\theta^{2}_{m}$ & $\eta^{2}_{m}$ \\
\addlinespace[0.125em]
\midrule
$F_{1}$  & 0.0091  & 5.11$\times10^{8}$  & 1.40$\times10^{1}$ \\
$F_{2}$  & 0.2301  & 2.94$\times10^{4}$  & 2.03$\times10^{-2}$ \\
$F_{3}$  & 0.7572  & 1.39$\times10^{3}$  & 3.17$\times10^{-3}$ \\
$F_{4}$  & 1.51    & 1.70$\times10^{3}$  & 7.67$\times10^{-3}$ \\
$F_{5}$  & 2.18    & 9.73$\times10^{1}$  & 6.36$\times10^{-4}$ \\
$F_{6}$  & 3.30    & 5.21$\times10^{2}$  & 5.15$\times10^{-3}$ \\
$F_{7}$  & 4.37    & 1.45$\times10^{1}$  & 1.90$\times10^{-4}$ \\
$F_{8}$  & 5.83    & 1.90$\times10^{2}$  & 3.32$\times10^{-3}$ \\
$F_{9}$  & 7.29    & 2.82$\times10^{0}$  & 6.16$\times10^{-5}$ \\
$F_{10}$ & 9.11    & 8.30$\times10^{1}$  & 2.27$\times10^{-3}$ \\
$F_{11}$ & 10.99   & 9.15$\times10^{-1}$ & 3.02$\times10^{-5}$ \\
$F_{12}$ & 13.14   & 4.09$\times10^{1}$  & 1.61$\times10^{-3}$ \\
$F_{13}$ & 15.39   & 3.22$\times10^{-1}$ & 1.48$\times10^{-5}$ \\
$F_{14}$ & 17.90   & 2.25$\times10^{1}$  & 1.21$\times10^{-3}$ \\
$F_{15}$ & 20.54   & 1.10$\times10^{-1}$ & 6.79$\times10^{-6}$ \\
$F_{16}$ & 23.40   & 1.35$\times10^{1}$  & 9.44$\times10^{-4}$ \\
$F_{17}$ & 26.39   & 6.12$\times10^{-2}$ & 4.85$\times10^{-6}$ \\
$F_{18}$ & 29.59   & 8.40$\times10^{0}$  & 7.46$\times10^{-4}$ \\
$F_{19}$ & 32.96   & 2.10$\times10^{-2}$ & 2.08$\times10^{-6}$ \\
$F_{20}$ & 36.48   & 5.40$\times10^{0}$  & 5.90$\times10^{-4}$ \\
$F_{21}$ & 40.15   & 1.17$\times10^{-2}$ & 1.41$\times10^{-6}$ \\
$F_{22}$ & 44.02   & 3.73$\times10^{0}$  & 4.92$\times10^{-4}$ \\
$F_{23}$ & 48.08   & 4.75$\times10^{-3}$ & 6.85$\times10^{-7}$ \\
$F_{24}$ & 52.26   & 2.62$\times10^{0}$  & 4.11$\times10^{-4}$ \\
$F_{25}$ & 56.57   & 2.15$\times10^{-3}$ & 3.65$\times10^{-7}$ \\
$F_{26}$ & 61.04   & 1.93$\times10^{0}$  & 3.53$\times10^{-4}$ \\
$F_{27}$ & 65.75   & 5.55$\times10^{-4}$ & 1.10$\times10^{-7}$ \\
$F_{28}$ & 70.55   & 1.45$\times10^{0}$  & 3.06$\times10^{-4}$ \\
$F_{29}$ & 75.48   & 3.13$\times10^{-4}$ & 7.10$\times10^{-8}$ \\
$F_{30}$ & 80.51   & 1.10$\times10^{0}$  & 2.65$\times10^{-4}$ \\
$F_{31}$ & 85.79   & 5.51$\times10^{-6}$ & 1.42$\times10^{-9}$ \\
$F_{32}$ & 91.17   & 8.53$\times10^{-1}$ & 2.33$\times10^{-4}$ \\
$F_{33}$ & 96.67   & 9.49$\times10^{-6}$ & 2.75$\times10^{-9}$ \\
$F_{34}$ & 102.26  & 6.65$\times10^{-1}$ & 2.04$\times10^{-4}$ \\
$F_{35}$ & 108.06  & 1.62$\times10^{-4}$ & 5.26$\times10^{-8}$ \\
$F_{36}$ & 113.96  & 5.30$\times10^{-1}$ & 1.81$\times10^{-4}$ \\
$F_{37}$ & 119.98  & 3.57$\times10^{-4}$ & 1.29$\times10^{-7}$ \\
$F_{38}$ & 126.07  & 4.29$\times10^{-1}$ & 1.62$\times10^{-4}$ \\
$F_{39}$ & 132.38  & 3.68$\times10^{-4}$ & 1.46$\times10^{-7}$ \\
$F_{40}$ & 138.76  & 3.50$\times10^{-1}$ & 1.46$\times10^{-4}$ \\
\toprule
$L_{1}$  & 0.6953   & 2.97$\times10^{0}$     & 6.19$\times10^{-6}$ \\
$L_{2}$  & 4.49     & 2.15$\times10^{2}$     & 2.89$\times10^{-3}$ \\
$L_{3}$  & 26.73    & 6.94$\times10^{-6}$    & 5.57$\times10^{-10}$ \\
$L_{4}$  & 32.06    & 4.18$\times10^{-1}$    & 4.02$\times10^{-5}$ \\
$L_{5}$  & 34.38    & 1.94$\times10^{-5}$    & 2.01$\times10^{-9}$ \\
$L_{6}$  & 46.09    & 4.36$\times10^{-6}$    & 6.03$\times10^{-10}$ \\
$L_{7}$  & 46.39    & 2.66$\times10^{-1}$    & 3.70$\times10^{-5}$ \\
$L_{8}$  & 54.72    & 4.21$\times10^{-6}$    & 6.91$\times10^{-10}$ \\
$L_{9}$  & 64.42    & 1.06$\times10^{-1}$    & 2.05$\times10^{-5}$ \\
$L_{10}$ & 66.04    & 2.89$\times10^{-6}$    & 5.72$\times10^{-10}$ \\
$L_{11}$ & 74.94    & 2.88$\times10^{-6}$    & 6.48$\times10^{-10}$ \\
$L_{12}$ & 80.14    & 8.36$\times10^{-2}$    & 2.01$\times10^{-5}$ \\
$L_{13}$ & 85.55    & 2.10$\times10^{-6}$    & 5.39$\times10^{-10}$ \\
$L_{14}$ & 95.09    & 6.15$\times10^{-6}$    & 1.75$\times10^{-9}$ \\
$L_{15}$ & 97.89    & 4.70$\times10^{-2}$    & 1.38$\times10^{-5}$ \\
$L_{16}$ & 106.02   & 1.38$\times10^{-6}$    & 4.39$\times10^{-10}$ \\
$L_{17}$ & 113.94   & 4.16$\times10^{-2}$    & 1.42$\times10^{-5}$ \\
$L_{18}$ & 115.63   & 4.59$\times10^{-6}$    & 1.59$\times10^{-9}$ \\
$L_{19}$ & 126.12   & 3.98$\times10^{-6}$    & 1.51$\times10^{-9}$ \\
$L_{20}$ & 130.98   & 2.77$\times10^{-2}$    & 1.09$\times10^{-5}$ \\
$L_{21}$ & 136.13   & 1.06$\times10^{-6}$    & 4.33$\times10^{-10}$ \\
\hline \hline
\end{tabular}
\caption{Extracted parameters $\theta^{2}_{m}$ and $\eta^{2}_{m}$ for the geometry with DBRs.}
\label{tab:SI_couplings_with}
\end{minipage}
\end{table}

\end{document}